\documentclass[prd,11pt,tightenlines,twoside,nofootinbib,showpacs]{revtex4}
\usepackage{graphicx,latexsym,mathrsfs,amsmath,amssymb,amsthm,amsopn,mathbbol,bm} 
\usepackage[dvips]{color}

\newenvironment{Equation*}{\equation}
{\endequation}
\newenvironment{Eqnarray*}{\eqnarray}
{\endeqnarray}

\newcommand{\La}[1]{\Lambda^{(#1)}} 
\newcommand{\Lv}[1]{\Lambda^{(#1;\text{vac})}}
\newcommand{\G}[1]{\Gamma^{(#1)}} 
\newcommand{\Gv}[1]{\Gamma^{(#1;\text{vac})}}
\newcommand{\Gt}{G^{(2)}}
\newcommand{\Gtv}{G^{(2,\text{vac})}}
\newcommand{\TP}{\Theta}
\newcommand{\I}{[I]}
\newcommand{\II}{[II]}
\newcommand{\ii}{\ensuremath{\mathrm{i}}}
\DeclareMathOperator{\Tr}{Tr}
\newcommand{\funcint}[2]{\int_{\mathcal{C}} \text{d}({#2}) \; #1}
\newcommand{\funcd}[2]{\frac{\delta #1}{\delta #2}}
\newcommand{\D}{\ensuremath{\mathrm{D}}}
\newcommand{\op}[1]{\ensuremath{\bm{\mathrm{#1}}}}
\newcommand{\erw}[1]{\ensuremath { %
    \left \langle {#1} \right \rangle}}
\newcommand{\Lag}{\ensuremath{\mathscr{L}}}
\newcommand{\feynint}[1]{\ensuremath{\int \frac{\mathrm{d}^{d} {#1}}{(2
\pi)^{d}}}}
\newcommand{\fint}[1]{\ensuremath{\int \frac{\mathrm{d}^4 #1}{(2\pi)^4}}}
\renewcommand{\d}{\ensuremath{\mathrm{d}}}
\DeclareMathOperator{\im}{Im}
\DeclareMathOperator{\re}{Re}
\newcommand{\R}{\ensuremath{\mathbb{R}}}
\DeclareMathOperator{\artanh}{artanh}

\begin{document}

\title{Renormalization in Self-Consistent Approximation schemes \\ at
  Finite Temperature III: \\ Global Symmetries} 
\author{Hendrik van Hees}
\affiliation{Fakult{\"at} f{\"u}r Physik, Universit{\"a}t Bielefeld,
  Universit{\"a}tsstra{\ss}e, D-33615 Bielefeld} 
\author{J{\"o}rn Knoll}
\affiliation{GSI Darmstadt, Planckstra{\ss}e 1, D-64291 Darmstadt}
\date{May 08, 2002}

\begin{abstract}
  
  We investigate the symmetry properties for Baym's $\Phi$-derivable
  schemes. We show that in general the solutions of the dynamical equations
  of motion, derived from approximations of the $\Phi$-functional, do not
  fulfill the Ward-Takahashi identities of the symmetry of the
  underlying classical action, although the conservation laws for
  the expectation values of the corresponding Noether currents are
  fulfilled exactly for the approximation. Further we prove that one can
  define an effective action functional in terms of the self-consistent
  propagators which is invariant under the operation of the same symmetry
  group representation as the classical action.  The requirements for this
  theorem to hold true are the same as for perturbative approximations: The
  symmetry has to be realized linearly on the fields and it must be free of
  anomalies, i.e., there should exist a symmetry conserving regularization
  scheme. In addition, if the theory is renormalizable in Dyson's narrow
  sense, it can be renormalized with counter terms which do not violate the
  symmetry.

\end{abstract}


\maketitle

\section{Introduction}

Symmetry principles are of fundamental importance in physics. For instance
the standard model of elementary particle physics rests on the principle of
local gauge symmetries.  These symmetries do not only guarantee the
observed conservation laws but also ensure the physical consistency of the
quantum field theory.  Especially no unphysical degrees of freedom for the
vector particles appear falsely in the particle spectrum. This ensures the
causality and the positive definiteness of probabilities (in the vacuum
case especially the unitarity of the S-Matrix).

Though fulfilled for the exact theory such symmetry concepts are commonly
violated in most of the approximation schemes, in particular if partial
resummations are performed. In this paper we address the question of global
symmetries for Dyson resummation schemes based on the two-particle
irreducible (2PI) action formalism \cite{lw60,leeyang61,cjt74} also known
as $\Phi$-derivable approximations \cite{baym62}. Here $\Phi$ denotes the
2PI part of the generating functional $\Gamma$. As Baym could show such
approximations have the property that expectation values of the Noether
currents are exactly conserved and the solutions are thermodynamically
consistent.  These properties even survive a consistent gradient
approximation of the Kadanoff-Baym equations leading to generalized quantum
kinetic equations \cite{KIV01} which dynamically treat the full spectral
functions of the particles and therefore permit a consistent transport
treatment of resonances.

In the first paper of this series \cite{vHK2001-Ren-I}, in the following
referred to as {\I}, we could verify an other important property of this type
of approximations, namely that of renormalizability, provided the original
theory is renormalizable in the standard sense. Using the path integral
techniques \cite{cjt74} within the Schwinger-Keldysh real time formalism
(for details and the notation conventions used here see {\I}) closed equations
for the renormalized self-consistent quantities with a counter term
structure solely defined at the vacuum level could be established. First
examples which include the fully self-consistent approximation at second
order in the coupling, i.e., including the tadpole and the sunset self-energy
diagrams, were given in the second part of this series
\cite{vHK2001-Ren-II}, referred to as {\II}. 

While for the solution to $\Phi$-derivable approximations the expectation
values of Noether currents are exactly conserved, this is no longer
guaranteed at the correlator level or for higher order vertex functions.
Thus in general the Ward-Takahashi identities (and thus also the
symmetries of the underlying classical action) are violated already for the
self-consistent self-energy. This was seen by Baym and Grinstein
investigating various approximations for the O($N$)-model \cite{baymgrin}.
Considerations related to symmetries in approximation schemes were
conducted already very early, e.g., in investigations of superfluid Helium
within non-relativistic many-body theory\cite{hohmar65}. At that time the
notion of gap-less approximations was coined in the context of
Nambu-Goldstone modes in spontaneously broken phases. In some recent
attempts the same phenomena were investigated in \cite{len99} comparing
various regularization and renormalization techniques, while in
\cite{nem99} the equations of motion were modified by introducing
mean-field dependent effective vertex functions in the approximations at
the gap-equation level.

In this paper we like to investigate the symmetry aspects of approximation
schemes at a quite general level fully within the 2PI functional approach.
In Sect. \ref{sect-wti} we derive generalized Ward-Takahashi identities for
the 2PI action functional. This is done in close analogy to the
corresponding investigations for the usual perturbative quantum action
functional (see for instance \cite{jl64}).  Thereby we prove the following
properties of the 2PI action functional formalism: Provided the symmetry is
realized as a linear representation on the fields and is free of anomalies
the considerations show the following:
\begin{enumerate}
\item[(1)]{If the 2PI action is truncated in a systematic way (loop
    expansion, $1/N$ expansion) it fulfills the same symmetries as the
    classical action if both, the classical fields and the Green's
    functions, are simultaneously transformed as tensors of the symmetry
    representation of 1$^{\text{st}}$ and 2$^{\text{nd}}$ rank,
    respectively.}
\item[(2)]{For the solutions of the corresponding self-consistent Dyson
    equation:
\begin{enumerate}
\item[(a)] 
expectation values of Noether currents are conserved. 
\item[(b)] The scheme is void of double counting and fulfills detailed
    balance relations and is thus thermodynamically consistent.
  \item[(c)]{However, Ward-Takahashi identities at the correlator level and
      for higher order vertex functions can be violated.}
\end{enumerate}
}
\end{enumerate}
The latter problem is expected for such partial resummation schemes since from
the point of view of the exact theory, which fulfills the symmetry, this
expansion is of course incomplete.  Diagrammatically it is easy to see that in
general $\Phi$-functional based partial Dyson resummations already violate the
crossing symmetry of vertex functions in orders of the expansion parameter
beyond that chosen for the truncation of the $\Phi$-functional.

As an important result of this paper it will be shown in Sect.
\ref{sect-eff-act} how to overcome these problems. The strategy is that for
any $\Phi$-derivable approximation based on a truncated 2PI action
$\Gamma[\varphi,G]$ one always can construct a non-perturbative
approximation for the effective quantum action on top of the
self-consistent solution which indeed recovers the symmetry in the above
cited sense. This effective action, called $\tilde{\Gamma}[\varphi]$ is
symmetric. It is expressed in terms of self-consistent propagators
$\tilde{G}[\varphi]$ at a given mean field and provides a non-perturbative
approximation to the usual 1PI quantum-action functional.  This action was
already considered in ref. \cite{cjt74} where the identity of vertex
functions derived from $\Gamma$ and $\tilde{\Gamma}$ has been shown for the
exact case. Here we concentrate on the consequences of truncation schemes,
where both pictures are no longer equivalent. Rather in terms of
perturbative diagrams, the new action indeed supplements that minimal set
of extra diagrams needed to recover the symmetry for the restricted set of
diagrams resummed by the $\Phi$-derivable Dyson scheme.  The extra terms
are encoded in a special Bethe-Salpeter equation and higher vertex
equations which precisely cope with the chosen $\Phi$-derivable scheme.
Formally the functional derivatives of this effective action with respect
to the background field, taken at the stationary point, define
non-perturbative approximations to the self-energies and proper vertex
functions which fulfill the usual Ward-Takahashi identities. We call the so
generated vertex functions \emph{external}, as they do not take part in the
self-consistent scheme but are rather constructed a posteriori, once the
self-consistent Dyson solutions are given. In this way symmetry-conserving
non-perturbative approximations to these quantities are provided. It is
shown that this also holds true for the renormalized physical quantities.
Since the external vertex functions are defined as usual by multiple
functional derivatives of the effective action all symmetries inherent in
this functional like crossing symmetry are guaranteed for the solutions.
We shall concentrate on the properties of the self-energy, which, if
derived from $\tilde{\Gamma}$, respects the symmetries (e.g., posses
Nambu-Goldstone modes).  This general scheme, however, permits to repair
the symmetry of any other higher order vertex functions such as the
correlators of Noether currents $\left<j_{\mu}(x)j_{\nu}(y)\right>$, which
then are conserved in contrast to the ones given by the underlying
self-consistent Dyson resummation.

In Sect. \ref{sect-on-funct} we discuss different approximation levels for
the 2PI action applied to the O($N$) linear $\sigma$-model.  Besides the
Hartree approximation (truncating the functional at the order $\lambda$) we
also discuss the next approximation level, i.e., up to second order in the
coupling. As the most simple example for the recovery of the
O($N$)-symmetry violated by the Hartree approximation in the broken
Nambu-Goldstone phase we show that the well known Random-Phase
approximation (RPA), see, for instance, \cite{aou96}, is derivable from a
non-perturbative effective action of the here defined kind. Some explicit
numerical results are given for the chiral linear $\sigma$-model. This
clarifies the question of symmetry violations of the Hartree approximation
shown in \cite{baymgrin} from the general point of view elaborated in this
paper, which specifies the route to symmetry preserving vertex functions
for any kind of $\Phi$-derivable approximation. In the context of this
series it is shown in sect. \ref{sect-ren-bs} that a straight forward
generalization of the renormalization procedure presented in the first part
of this series {\I} also leads to \emph{closed renormalized} equations for
these vertex functions at finite temperature which counter-term structures
are solely defined on the vacuum level. The paper is closed with conclusion
and outlook.

The appendices contain a formal derivation of the expression for the
$\Phi$-functional which uses the path-integral formalism developed in
\cite{cjt74} and a short summary about Noether's theorem in the
classical context which is needed in Sect. \ref{sect-wti} to formulate
the generalized Ward-Takahashi identities.

\section{Symmetry properties of the exact 2PI action functional}
\label{sect-wti}

The 2PI action functional \cite{lw60,leeyang61,baym62} can be introduced
with help of the well known path integral methods \cite{cjt74}. Besides the
standard local source $J$ one also supplements a bilocal source $B$ to the
partition sum $Z=Z[J,B]$.  The 2PI action functional $\Gamma[\varphi,G]$ is
then defined as the double Legendre transformation of $W[J,B]=-\ii \hbar
\ln Z[J,B]$ with respect to both, $J$ and $B$. Details and a slightly
different derivation than in \cite{cjt74} are given in Appendix
\ref{sect-baym-funct}.  A formal loop expansion of the path integral yields
\begin{equation}
\label{3n.5}
\Gamma[\varphi,G] = S[\varphi] + \frac{\ii \hbar}{2} \Tr \ln (G_{12}^{-1}/M^{2})
  +\frac{\ii \hbar}{2}
  \funcint{\mathscr{D}_{12}^{-1}(G_{12}-\mathscr{D}_{12})}{12} +
  \Phi[\varphi,G].
\end{equation}
Here we have used the notation $\funcint{f_{1}}{1}$ for the integration
over $d$-dimensional space time in the sense of dimensional regularization
and appropriate sums for the internal field components (i.e., ``charge
space'' indices). Within the real-time formalism adapted to the equilibrium
case the time integration runs along a modified Schwinger-Keldysh
contour\footnote{besides the real-time parts along the real time axis from
  an initial time $t_i$ to a final time $t_f$ and back to $t_i$ this
  contour also includes a branch parallel to the imaginary time axis from
  $t_{i}$ to $t_{i}-\ii \beta$ ($\beta=1/T$ denotes the inverse temperature
  of the system).  Finally we let the times $t_{i}$ and $t_{f}$ go to
  $-\infty$ and $+\infty$, respectively. All considerations can be extended
  to more general off-equilibrium initial statistical operators with the
  qualification that in this case it does not make sense to take $t_{i}
  \rightarrow -\infty$} which includes an imaginary branch of length
$\ii\beta$. In (\ref{3n.5}) $S[\varphi]$ denotes the classical action as a
functional of the mean field $\varphi$, while the terms proportional to
$\hbar$ account for the quantum fluctuations at the one loop level. The
free propagator in the presence of the mean field $\varphi$ is denoted by
$\mathscr{D}$:
\begin{equation}
\label{3n.6}
(\mathscr{D}^{-1})_{12}=\frac{\delta^{2} S[\varphi]}{\delta \varphi_{1}
  \delta \varphi_{2}}.
\end{equation}
Finally $\Phi$ accounts for all higher order terms. It is diagrammatically
defined as the sum of all \emph{two-particle irreducible (2PI) closed
  diagrams} beyond the one-loop level which can be built with
point-vertices defined by the interaction part of $S[\phi+\varphi]$, i.e.,
the part with at least three $\phi$-fields expanded in the same way as the
Feynman diagrams of perturbation theory. Thereby lines stand for \emph{full
  propagators} $G$ rather than free propagators.  We shall give an
analytical definition of the $\Phi$ functional which can be used for
practical calculations of approximations like the coupling constant or loop
expansion for the $\Phi$-functional in appendix \ref{sect-baym-funct}.

The equations of motion for the mean field $\varphi$ and the Green's
function are given by the stationarity of $\Gamma$, i.e., at vanishing
auxiliary sources $J=B=0$ as
\begin{equation}
\label{3n.7}
  \funcd{\Gamma[\varphi,G]}{\varphi} 
  \stackrel{!}{=} 0,\quad
  \funcd{\Gamma[\varphi,G]}{G} 
  \stackrel{!}{=} 0.
\end{equation}
Using Eq. (\ref{3n.5}) the second equation is seen as the Dyson
equation for the full propagator
\begin{equation}
\label{3n.8}
(\mathscr{D}^{-1})_{12} - (G^{-1})_{12} = 2 \ii
  \funcd{\Phi[\varphi,G]}{G_{12}} := \Sigma_{12}.
\end{equation}
Thus (\ref{3n.7}) gives a closed self-consistent set of equations of motion
for the mean fields and the self-energies in terms of the exact Green's
function $G$. The 2PI property of $\Phi$ avoids double counting in the
sense that the diagrams of the self-energy with lines denoting exact
Green's functions do not contain any self-energy insertion in any of its
lines, i.e., it generates \emph{one-particle irreducible (1PI) skeleton
  diagrams} for the self-energy by variation with respect to $G$. This is
immediately clear from the fact that the derivative of $\Phi$ with respect
to $G$ diagrammatically implies to open anyone of its lines in the diagrams
building $\Phi$ and taking the sum over the so obtained diagrams with two
truncated external points.

For the discussion of the symmetry properties of the above defined
functionals we take the O($N$)-symmetric $\phi^4$-theory as an example. The
generalization to other models and symmetries with more complicated field
configurations is straight forward.

We use the fact that the path integral measure is invariant under a field
translation $\vec{\phi}'=\vec{\phi}+\delta \vec{\phi}$. In first order of
$\delta \vec{\phi}$ we find
\begin{equation}
\label{3.1}
\begin{split}
  0= & \int \D \vec{\phi} \left [\funcint{\left (\funcd{S}{\phi_{1}^{j}} +
        J_{1j} \right )\delta \phi_{1}^{j}}{1} + \funcint{B_{j1,k2}
      \phi_{1}^{j} \delta \phi_{2}^{k}}{12} \right] \\
  & \times \exp \left [ \ii S[\vec{\phi}] + \ii \funcint{J_{1j}
      \phi_{1}^{j}}{1} + \frac{\ii}{2} \funcint{B_{j1,k2} \phi_{1}^{j}
      \phi_{2}^{k}}{12} \right].
\end{split}
\end{equation}
Here subscripts and superscripts $j,k$ denote the field components
  $(\phi^j)=\vec{\phi}$, and Einstein's summation convention is implied.
For a \emph{local} O($N$)-transformation
\begin{equation}
\label{3.2}
\delta \phi_{1}^{j}=\ii \delta \chi_{1}^{a} {(\tau^{a})^{j}}_{j'}
\phi_{1}^{j'} 
\end{equation}
we obtain
\begin{equation}
\label{3.3}
\begin{split}
  0 = & \int \D \phi \int_{\mathcal{C}} \d(1) \left [ \left
      (\funcd{S}{\phi_{1}^{j}} +J_{1j} \right ) \ii {(\tau^{a})^{j}}_{j'}
    \phi_{1}^{j'} \delta \chi_{1}^{a} + 2 \funcint{B_{j1,k2} \ii
      {(\tau^{a})^{j}}_{j'} \phi_{1}^{j'}
      \phi_{2}^{k}}{2} \delta \chi_{1}^{a} \right] \\
  & \times \exp \left[\ii S[\vec{\phi}] + \ii \funcint{J_{1'j'}
      \phi_{1'}^{j'}}{1'} + \frac{\ii}{2} \funcint{B_{1'j',2'k'}
      \phi_{1'}^{j'} \phi_{2'}^{k'}}{1'2'} \right].
\end{split}
\end{equation}
Since the classical action functional is invariant under \emph{global}
O($N$) transformations (\ref{3.2}), c.f. Appendix \ref{sect-baym-funct}, we
read off from (\ref{a.8}) that (\ref{3.3}) can be expressed in the form
\begin{equation}
\label{3.4}
\begin{split}
  0 = & \funcint{\delta \chi_{1}^{a} \left \{ \partial_{\mu} j_{1}^{a \mu}
      \left [\frac{\delta}{\ii \delta \vec{J}} \right ] Z[\vec{J},B] +
      J_{1j} \ii {(\tau^{a})^{j}}_{j'} \funcd{Z[\vec{J},B]}{\ii J_{1j'}}
    \right \}}{1} \\ & + 2 \funcint{\delta \chi_{1}^{a} B_{1j,2k} \ii
    {(\tau^{a})^{j}}_{j'} \funcd{Z[\vec{J},B]}{\ii B_{1j',2k}}}{12}.
\end{split}
\end{equation}
Since $\delta \chi_{1}^i$ is an arbitrary function this can be brought to
the local expression
\begin{equation}
\label{3.5}
0 =  \partial_{\mu}
  j_{1}^{a \mu} \left [\frac{\delta}{\ii \delta \vec{J}} \right ] Z[\vec{J},B] +
J_{1j} \ii {(\tau^{a})^{j}}_{j'} \funcd{Z[\vec{J},B]}{\ii
  J_{1j'}} + 2 \funcint{B_{1j,2k} \ii {(\tau^{a})^{j}}_{j'} 
\funcd{Z[\vec{J},B]}{\ii B_{1j',2k}}}{2}.
\end{equation}
For the solution of the equations of motion we have $\vec{J}=0$ and $B=0$.
Thus the expectation value of the current is conserved as we read off from
(\ref{3.5}):
\begin{equation}
\label{3.6}
\partial_{\mu}
  j_{1}^{a \mu} \left [\frac{\delta}{\ii \delta \vec{J}} \right ]
  Z[\vec{J},B] = Z[\vec{J},B] \partial_{\mu} \erw{\op{j}_{1}^{a \mu}}=0.
\end{equation}
Here the expectation value has to be read as the quantum statistical
expectation value for the local current operator if interpreted within the
operator formalism of quantum field theory. Here and in the following we
write operators as bold face upright symbols.

For sake of completeness we write down the local Ward-Takahashi identities
(WTIs) for the current. From (\ref{a.11}) we find for the O($N$)-Noether
current the expression
\begin{equation}
\label{3.7}
j_{1\mu}^{a}[\phi]=-\ii (\tau^{a})_{jj'} \phi_{1}^{j} \partial_{\mu}
\phi_{1}^{j'}.
\end{equation}
Using this in (\ref{3.5}) yields the local WTIs for the functional $W=-\ii
\ln Z$:
\begin{equation}
\label{3.8}
-2 \left[ \Box^{(2)} \funcd{W}{B_{1k,2k'}} \right]_{1=2}
(\tau^{a})_{kk'} + \vec{J}_{1j} {({\tau}^{a})^{j}}_{j'} 
\funcd{W}{J_{1j'}} + 2
\funcint{B_{j1,k2} {(\tau^{a})^{j}}_{j'} \funcd{W}{B_{j'1,k2}}}{2}=0.
\end{equation}
Finally we also obtain an expression for the $\Gamma$-functional by using
the relations (\ref{2.5}) and (\ref{2n.15}):
\begin{equation}
\label{3.9}
\begin{split}
  -\ii \partial_{\mu} \erw{\op{j}_{1}^{\mu}[\op{\phi}]} & = \left
    (-\varphi_{1}^{k} \Box \varphi_{1}^{k'} + 2\ii \left [ \Box^{(2)}
      G_{1k,2k'} \right]_{1=2} \right ) (\tau^{a})_{kk'} \\ & =
  \funcd{\Gamma}{\vec{\varphi}_{1}} \hat{\tau}^{a} \vec{\varphi}_{1} + 2
  \funcint{\funcd{\Gamma}{G_{j1,k2}} \tau_{jj'}^{a} G_{1j',2k}}{2}.
\end{split}
\end{equation}
Again it follows immediately that for the solutions of the equations of
motion (\ref{3n.7}) the expectation value of the O($N$)-Noether current is
conserved. Taking further derivatives of (\ref{3.5}) yields WTIs for higher
order Green's functions of the Noether current, which we do not need in the
further line of arguments.

For the further symmetry analysis we like to concentrate on properties of
the 2PI functional (\ref{3n.5}). We thus go back to (\ref{3.3}) and set
$\delta \chi^{i}=\text{const}$. Then the first term in the curly bracket is
a complete divergence and thus the integral vanishes and due to the linear
independence of the constants $\delta \chi^i$ we obtain the \emph{global
  form} of the generalized \emph{Ward-Takahashi identities}.

Especially this proves that the generating functional $Z$ is invariant
under \emph{global} O($N$) transformations when the local source $J_{1j}$ is
transformed in the contragredient way of the fields and the bilocal sources
$B_{1j,2k}$ as a covariant tensor of 2$^{\text{nd}}$ rank.  Since
(\ref{3.4}) is of first order in the derivatives of $Z$ with respect to the
sources the same holds true for $W$:
\begin{equation}
\label{3.10}
\funcint{J_{1j} \ii {(\tau^{a})^{j}}_{j'}
  \funcd{W[\vec{J},B]}{J_{1j'}}}{1} +
\funcint{ \left [ B_{1j',2k} {(\tau^{a})^{j'}}_{j} +  
    B_{1j,2k'} {(\tau^{a})^{k'}}_{k} \right ]
  \funcd{W[\vec{J},B]}{B_{j1,k2}}}{12},
\end{equation}
and from (\ref{2.5}) we find (again for \emph{global} transformations):
\begin{equation}
\label{3.11}
\funcint{\funcd{\Gamma[\vec{\varphi},G]}{\vec{\varphi}_{1}}
  \hat{\tau}^{a} \vec{\varphi}_{1}}{1} +
\funcint{\funcd{\Gamma[\vec{\varphi},G]}{G_{12}^{jk}} \left
    [ {(\tau^{a})^{j}}_{j'} G_{12}^{j'k} + {(\tau^{a})^{k}}_{k'}
    G_{12}^{jk'} \right ]}{12}=0. 
\end{equation}
This result can be derived also directly from (\ref{3.9}) by taking the
integral $\funcint{\dots}{1}$ on both sides of the equation.

\section{Symmetries of 2PI $\Phi$-derivable approximations}
\label{sect-eff-act}

So far the Ward-identities were derived for the \emph{exact
    functionals, Green's functions and mean fields}. In particular it
  is important to realize that the identity
\begin{equation}
\label{4.1}
\ii G_{12}^{jk} :=-\ii \frac{\delta^{2}W[J,B]}{\delta J_{1j} \delta J_{2k}} =
2 \funcd{W[J,B]}{B_{1j,2k}}- \varphi_{1}^{j} \varphi_{2}^{k}
\text{ for the \emph{exact functional}}
\end{equation}
was proven using the underlying path integral definition of the exact
functionals. We now step towards the $\Phi$-derivable approximations.  They
are defined by a \emph{truncation} of the auxiliary functional
$\Phi[\varphi,G]$ in the definition (\ref{3n.5}) of the generating
functional $\Gamma[\varphi,G]$ while keeping the variational
properties (\ref{3n.7}) which define the equations of motion for mean field
and propagator
\begin{equation}
\label{3n.7-approx}
  \funcd{\Gamma_{\text{apprx}}[\varphi,G]}{\varphi} 
  \stackrel{!}{=} 0,\quad
  \funcd{\Gamma_{\text{apprx}}[\varphi,G]}{G} 
  \stackrel{!}{=} 0.
\end{equation}
The latter equation of motion defines the self-energy in terms of the
self-consistent propagator by
\begin{equation}
\Sigma_{12} := (\mathscr{D}^{-1})_{12} - (G^{-1})_{12} = 2 \ii
  \funcd{\Phi_{\text{apprx}}[\varphi,G]}{G_{12}}.
\end{equation}
Thereby $\Phi[\varphi,G]$ can be truncated according to various schemes,
like expansion in powers of $\lambda$, in loop order or in powers of $1/N$
(see \cite{ahr02}), excluding the internal structure of the Green's
functions from the counting. The sole requirement for the thus constructed
$\Phi_{\text{apprx}}$ is that it remains \emph{invariant} under the
symmetry transformations as explained in appendix \ref{sect-baym-funct},
i.e., that the generalized WTIs (\ref{3.11}) hold true also for the
approximation. In our case of the linear O($N$)-model this is the case for
the just mentioned schemes. The preferred choice may be motivated by the
physical problem.

Since in the so defined 2PI approximation one has a functional structure
solely defined in terms of the approximated one- and two-point functions
$\varphi$ and $G$, path integral properties and relations such as
(\ref{4.1}) do not need to hold true any longer. Thus irrespective of the
chosen truncation scheme the $n$-point functions defined by the
approximated 2PI functional generally may no longer coincide with the
corresponding 1PI vertex-functions and thus WTIs derived from the 1PI
formalism may be violated for those self-consistent approximations.
  
The reason is simply seen in diagrammatic terms. The Schwinger-Dyson
resummation creates a very restricted subset of 1PI diagrams resulting from
the iterative insertion of self-energy pieces.  In particular this implies
that already the crossing symmetry is violated at orders of the respective
expansion parameter beyond those included in $\Phi_{\text{apprx}}$.  Thus
it has to be expected that in general for the self-consistent approximation
scheme (\ref{3n.7-approx}) the WTIs for the vertex functions derived from
the 1PI formalism are not fulfilled for the self-consistent approximations
of the self-energy and higher $n$-point functions derived from the 2PI
functional. For example the symmetries may already be violated for the
self-energy $\Sigma$. Especially for systems in the Nambu-Goldstone phase
this implies that the self-consistent propagators may not comply with
Goldstone's theorem.

In order to cure this problem we supplement the 2PI approximation scheme
(\ref{3n.7-approx}) by an additional effective action
$\tilde{\Gamma}_{\text{apprx}}[\varphi]$ defined with respect to the
self-consistent solution as
\begin{equation}
\label{4.2}
\begin{split}
& \tilde{\Gamma}_{\text{apprx}}[\varphi]=
\Gamma_{\text{apprx}}[\varphi,\tilde{G}[\varphi]] \text{ with
  $\tilde{G}[\varphi]$ defined by} \\
& \left . \funcd{\Gamma_{\text{apprx}}[\varphi,G]}{G}
\right|_{G=\tilde{G}[\varphi]} 
\equiv 0.
\end{split}
\end{equation}
Here $\tilde{G}[\varphi]$ is given as the Schwinger-Dyson solution in
presence of a given mean field $\varphi$.  Strategies like this date back
to Baym and Kadanoff \cite{bk61}, the equivalence of both functionals
$\Gamma[\varphi,G]$ and $\tilde{\Gamma}[\varphi]$ at the exact level was
shown in ref.  \cite{cjt74}, while the consequences for truncation schemes
were discussed, e.g., in \cite{den96} in the context of a background field
formulation.  Here we will show that independent of the chosen trancation
scheme $\tilde{\Gamma}_{\text{apprx}}[\varphi]$ permits to construct proper
vertex functions on top of the self-consistent solutions of
(\ref{3n.7-approx}), which then obey the symmetries as described by the
usual 1PI WTIs. In diagrammatic terms, to any $\Phi$-derivable
approximation this procedure supplements that minimal set of
diagrams which is needed to recover the symmetries.  Effectively $\varphi$
can be considered as a background field and
$\tilde{\Gamma}_{\text{apprx}}[\varphi]$ permits to construct the linear
response of the system with respect to a fluctuation in $\varphi$ around
the self-consistent solution of the equations of motion
(\ref{3n.7-approx}). The latter coincides with the stationary point
$\tilde{\varphi}$ of the effective action functional
$\tilde{\Gamma}_{\text{apprx}}[\varphi]$
\begin{equation}
\label{4.3}
\begin{split}
  \left .
    \funcd{\tilde{\Gamma}_{\text{apprx}}[\vec{\varphi}]}{\vec{\varphi}}
  \right|_{\vec{\varphi}=\tilde{\varphi}} &= \left (
    \funcd{\Gamma_{\text{apprx}}[\vec{\varphi},G]}{\vec{\varphi}} +
    \funcint{\funcd{\Gamma_{\text{apprx}}[\vec{\varphi},G]}{G_{12}^{jk}}
      \funcd{\tilde{G}_{12}^{jk}[\vec{\varphi}]}{\vec{\varphi}}}{12}
  \right)_{G=\tilde{G}[\tilde{\varphi}],\vec{\varphi}=\tilde{\varphi}} \\
  & = \left (\funcd{\Gamma_{\text{apprx}}[\vec{\varphi},G]}{\vec{\varphi}}
  \right)_{G=\tilde{G}[\tilde{\varphi}],\vec{\varphi}=\tilde{\varphi}}
  \stackrel{!}{=} 0,
\end{split}
\end{equation}
i.e., $\tilde{\varphi}$ and $\tilde{G}[\tilde{\varphi}]$ solve
(\ref{3n.7-approx}).

Now since the derivation of the WTIs (\ref{3.11}) only relies on the
symmetry of the $\Gamma_{\text{apprx}}[\varphi,G]$-functional under
the O($N$)-transformations of $\varphi$ and $G$ as a contravariant
vector and a contravariant tensor of 2$^{\text{nd}}$ rank,
respectively, and this by construction holds true for the approximated
functional, relation (\ref{3.11}) is also valid for the approximation.
Therefore we immediately conclude that the \emph{non-perturbative
  effective action functional} (\ref{4.2}) is O($N$) symmetric in the
sense of the classical action, i.e.,
\begin{equation}
\label{4.4}
\funcint{\funcd{\tilde{\Gamma}_{\text{apprx}}[\vec{\varphi}]}{\varphi_{1}^{j}}
  {(\tau^{a})^{j}}_{j'} \varphi_{1}^{j'}}{1}=0.
\end{equation}
Thus, if we define self-energies and higher proper vertex functions by the
usual definition as multiple derivatives of $\tilde{\Gamma}$, then these
functions fulfill the \emph{usual} 1PI WTIs, and thus obey all symmetry
properties derived from them. We call the so constructed functions
\emph{external}, as they do not take part in the self-consistent scheme but
rather are calculated as a function of the frozen self-consistent solutions
of (\ref{3n.7-approx}), the latter being referred to as \emph{internal}.
While external and internal quantities coincide at the exact level, this is
commonly no longer the case for the here discussed approximation schemes
based on a truncated $\Phi$-functional.  Below we drop the label ``apprx''.

\subsection{Goldstone's Theorem}

As a first quantity we discuss the \emph{external propagator} defined
from $\tilde{\Gamma}$ by
\begin{equation}
\label{4.5}
(G_{\text{ext}}^{-1})_{1j,2k} = \left . \frac{\delta^{2}
    \tilde{\Gamma}[\varphi]}{\delta \varphi_{1}^{j} \delta
    \varphi_{2}^{k}} \right |_{\varphi=\tilde{\varphi}},
\end{equation}
which fulfills the WTI
\begin{equation}
\label{4.6}
\funcint{(G_{\text{ext}}^{-1})_{j1,k2} {(\tau^{a})^{j}}_{j'}
\varphi_{1}^{j'}}{1}=0,
\end{equation}
as can be immediately seen by taking the functional derivative of
(\ref{4.4}) with respect to $\varphi_{2}^{k}$ and using the fact that
$\tilde{\Gamma}$ is stationary for $\varphi=\tilde{\varphi}$. For a
translationally invariant situation (for instance for the vacuum or thermal
equilibrium) this can easily be Fourier transformed leading to
\begin{equation}
\label{4.7}
(G_{\text{ext}}^{-1})_{jk}(p=0){(\tau^{a})^{j}}_{j'} \varphi^{j'} =
-(M^2)_{jk} {(\tau^{a})^{j}}_{j'} \varphi^{j'}=0,
\end{equation} 
 where $M^{2}$ denotes the (thermal) mass matrix for the field degrees of
 freedom. This reflects Goldstone's theorem: There are as many massless
 states as group generators which do not annihilate the mean field, i.e.,
 the number of massless Goldstone bosons is the dimension of the symmetry
 group minus the dimension of the symmetry group of the mean field. In our
 case the symmetry group is O($N$) and the symmetry group of the mean field
 is O($N-1$), which means that we have $(N-1)N/2-(N-2)(N-1)/2=N-1$ massless
 Goldstone Bosons in the spontaneously broken symmetry phase.

\subsection{The external propagator}

Now we explicitly construct the external propagator.  From
(\ref{4.2}) we find
\begin{equation}
\label{4.8}
\left (G_{\text{ext}}^{-1} \right )_{1j,2k} = \left [ \frac{\delta^{2}
  \Gamma[\varphi,G]}{\delta \varphi_{1}^{j} \delta \varphi_{2}^{k}} +
  \funcint{\frac{\delta^{2} \Gamma[\varphi,G]}{\delta \varphi_{1}^{j}
  \delta G_{3'4'}^{j'k'}}
  \funcd{\tilde{G}_{3'4'}^{j'k'}}{\varphi_{2}^{k}}}{3'4'} \right
  ]_{G=\tilde{G}[\tilde{\varphi}],\varphi=\tilde{\varphi}}.
\end{equation}
The derivative of $\tilde{G}$ with respect to the background
field $\varphi$ can be expressed through the identity
\begin{equation}
\label{4.9}
\funcint{(\tilde{G}^{-1})_{1j,2'k'}
  \tilde{G}_{2'2}^{k'k}}{2'}=\delta_{12}^{(d)} \delta_{j}^{k}. 
\end{equation}
Taking its derivative with respect to the mean field yields
\begin{equation}
\label{4.10}
\funcint{ \left [ \funcd{(\tilde{G}^{-1})_{1j,2'k'}}{\varphi_{3}^{l}}
  \tilde{G}_{2'2}^{k'k} + (\tilde{G}^{-1})_{1j,2'k'}
  \funcd{\tilde{G}_{2'2}^{k'k}}{\varphi_{3}^{l}} \right ]}{2'} = 0, 
\end{equation}
implying
\begin{equation}
\label{4.11}
\funcd{G_{12}^{jk}}{\varphi_{3}^{l}} = -\funcint{
  \funcd{\tilde{G}^{-1}_{1'j',2'k'}}{\varphi_{3}^{l}} \tilde{G}_{11'}^{jj'}
  \tilde{G}_{22'}^{kk'}}{1'2'}.
\end{equation}
The functional derivative of $\tilde{G}^{-1}$, which is a three-point
function
\begin{equation}
\label{4.12}
\Lambda^{(3)}_{1j,2k;3l}=\funcd{\tilde{G}_{1j,2k}^{-1}}{\varphi_{3}^{l}},
\end{equation}
can be implicitly expressed in form of a \emph{Bethe-Salpeter
  equation} (BS)
\begin{equation}
\label{4.13}
\Lambda_{1j,2k;3l}^{(3)}=\Gamma_{1j,2k;3l}^{(3)} - \ii
  \funcint{\Gamma_{1j,2k;3'l',4'm'}^{(4)}
  \tilde{G}_{3'3''}^{l'l''} \tilde{G}_{4'4''}^{m'm''}
  \Lambda_{3''l'',4''m'';3l}^{(3)}}{3'4'3''4''}.
\end{equation}
We denote the resummed, i.e., two-particle reducible $n$-point vertex
functions by $\Lambda^{(n)}$ while the corresponding irreducible parts are
denoted by $\Gamma^{(n)}$. Constructed through $\Phi$ the latter
\begin{equation}
\label{4.14}
\begin{split}
  \Gamma_{1j,2k;3l}^{(3)} &= \left [\frac{\delta^{3} S[\varphi]}{\delta
      \varphi_{1}^{j} \delta \varphi_{2}^{k} \delta \varphi_{3}^{l}} - 2
    \ii \frac{\delta^{2} \Phi[\varphi,G]}{\delta G_{12}^{jk} \delta
      \varphi_{3}^{l}} \right
  ]_{G=\tilde{G}[\tilde{\varphi}], \varphi=\tilde{\varphi}}, \\
  \Gamma_{1j,2k;3l,4m}^{(4)} &= -2 \left [\frac{\delta^{2} \Phi[\varphi,G]}{\delta
      G_{12}^{jk} \delta G_{34}^{lm}}
  \right]_{G=\tilde{G}[\tilde{\varphi}], \varphi=\tilde{\varphi}}
\end{split}
\end{equation}
are functionals of the self-consistent propagator $\tilde{G}$. 

In terms of diagrams the BS-equation (\ref{4.13}) can be depicted as
follows:
\begin{equation}
\label{4.13b}
\ii \Lambda^{(3)}=\parbox{11.4cm}{\includegraphics{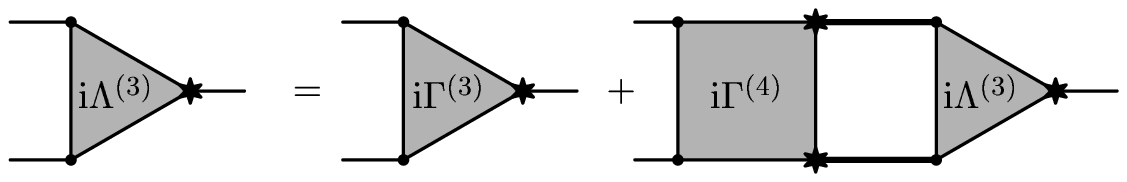}}.
\end{equation}
The different external points of the vertex functions, dots and stars,
indicate the different kind of points, which are separated by a semicolon
in (\ref{4.13}). This specialty of those diagrams clearly shows that only
certain channels are resummed namely those which are not contained in the
self-consistent scheme. As explained in {\I} the Dyson resummation implies
$s$-channel ladder resummations, while the here given BS-equation creates
$t$- and $u$-channel ladders, this way recovering crossing and all global
symmetries. This again shows the virtue of the $\Phi$-functional method.
Through the effective action functional $\tilde{\Gamma}[\varphi]$ it
permits the construction of fully symmetry preserving vertex functions void
of double counting, although each dynamical equation, the Dyson and
BS-equation, for themselves violate these symmetries. The kernel
$\Gamma^{(4)}$ of the BS-equation is symmetric with respect to the
simultaneous change of the pairs 1,2 with 3,4 and 2PI for separating these
pairs.  In addition the $\Phi$-functional ensures that especially the
crossing symmetry is also recovered for the counter terms, a property
necessary to renormalize the proper $n$-point functions.\footnote{Power
  counting and Weinberg's convergence theorem show that in our case the
  superficial degree of divergence of a diagram $\gamma$ is
  $\delta(\gamma)=4-E$, where $E$ is the number of external legs of the
  diagram. This means that due to the BPHZ-renormalization procedure only
  proper $n$-point functions with $n \leq 4$ have to be renormalized.}
This is non-trivial since the counter terms appear to arbitrary order in
the sense of an $\hbar$- or coupling constant expansion.

Using (\ref{4.13}), (\ref{4.14}), (\ref{2n.15}), and (\ref{2.4}) in
(\ref{4.8}) we finally obtain for the external self-energy
\begin{equation}
\label{4.15}
\begin{split}
  (\Sigma_{\text{ext}})_{1j,2k}=-\Bigg [ & \frac{\ii}{2}
  \funcint{\frac{\delta^{4}S[\varphi]}{\delta \varphi_{1}^{j} \delta
      \varphi_{2}^{k} \delta \varphi_{1'}^{j'} \delta \varphi_{2'}^{k'}}
    G_{1'2'}^{j'k'}}{1'2'} + \frac{\delta^{2} \Phi[\varphi,G]}{\delta
    \varphi_{1}^{j} \delta \varphi_{2}^{k}} \\
  & - \frac{\ii}{2} \funcint{\Gamma_{3'j',4'k';1j}^{(3)} G_{3'3''}^{j'j''}
    G_{4'4''}^{k'k''} \Lambda_{3''j'',4''k'';2k}^{(3)}}{3'4'3''4''} \Bigg
  ]_{G=\tilde{G}[\tilde{\varphi}],\varphi=\tilde{\varphi}}
\end{split}
\end{equation}
Again this can be easily expressed with help of diagrams
\begin{equation}
\label{4.16}
\raisebox{-3mm}{\parbox{12.6cm}{\includegraphics{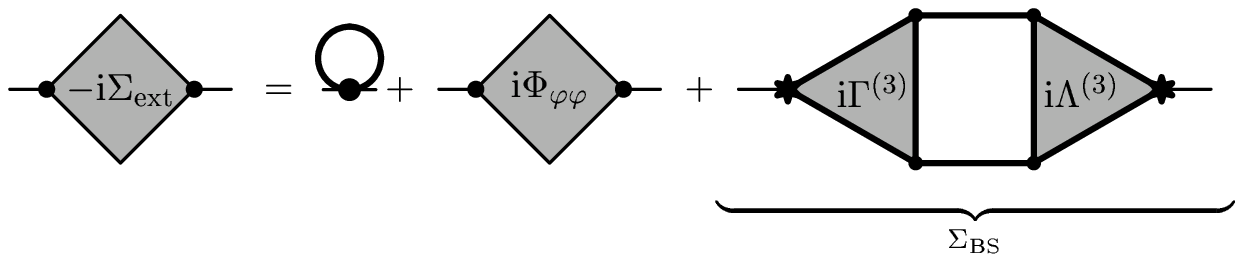}}},
\end{equation}
where the last term of the external self-energy, the BS-part
$\Sigma_{\text{BS}}$, needs special care with respect to renormalization.
The external self-energy $\Sigma_{\text{ext}}$, here derived for a general
$\Phi$-derivable truncation scheme is frequently called the mass matrix, as
it is used to determine the masses of the fluctuations in the phase of
broken symmetry.

\subsection{Solution and renormalization of the
  Bethe-Salpeter equation}
\label{sect-ren-bs}

The BS-equation now given in short hand four-point function notation
\begin{equation}
\label{4.13-short}
\Lambda^{(3)}=\Gamma^{(3)} + \Gamma^{(4)} G ^{(2)}\Lambda^{(3)},
\text{ where } G_{12;34}^{(2)}=-\ii G_{13} G_{24}
\end{equation}
can be solved in terms of the four-point function $\Lambda^{(4)}$ as
\begin{equation}
\label{4.13-L4}
\Lambda^{(3)} = \Gamma^{(3)} + \Lambda^{(4)} G^{(2)}\Gamma^{(3)},
\end{equation}
where the four-point function $\Lambda^{(4)}$ is defined through two
  equivalent BS-equations
\begin{alignat}{2}
  \Lambda^{(4)}&=\Gamma^{(4)} + \Gamma^{(4)} G^{(2)}\Lambda^{(4)}
  \label{BS-r} \\
  &= \Gamma^{(4)} + \Lambda^{(4)} G^{(2)}\Gamma^{(4)} \label{BS-l}.
\end{alignat}
The latter four-point BS-equation is identical to that already considered
in the first paper {\I} of this series, however with two differences: In I
only $s$-channel resummations were considered which restricts the momentum
arguments to forward scattering, and secondly the BS-equation was used
solely at the vacuum level.  Here, however, the BS-equation truly acts at
finite temperature. The latter implies that for the renormalization
procedure vacuum and finite temperature pieces need to be separated along
the lines given in {\I}. Again however the only subdiagrams to be renormalized
are the four-point functions which lead to the renormalized vacuum function
$\Lambda$ defined in {\I}.

Considering the difference between the four-point BS-equation at finite $T$
and that at vacuum leads to the following expression\footnote{As explained
  in detail in {\I}, in this subtraction technique all vacuum functions are
  contour diagonal, i.e., they vanish for arguments with mixed vertex
  placement}
\begin{equation}
\begin{split}
  \La{4}-\Lv{4} &= \G{4}-\Gv{4} + \G{4} \Gt \La{4} -\Gv{4} \Gtv \Lv{4} \\
  &= \G{4}-\Gv{4} + \La{4} \Gt (\G{4}-\Gv{4}) \\ & \quad + \La{4} \Gt \Gv{4} - \Gv{4}
  \Gtv
  \Lv{4} \\
  &= \G{4}-\Gv{4} + \La{4} \Gt (\G{4}-\Gv{4}) \\ & \quad + (\G{4}-\Gv{4})
  \Gtv \Lv{4} \\
  & \quad + \La{4} \Gt (\underbrace{\Lv{4}-\Gv{4} \Gtv \Lv{4}}_{\Gv{4}}) \\
  & \quad - (\underbrace{\La{4}-\La{4} \Gt \G{4}}_{\G{4}}) \Gtv \Lv{4} \\
  &=  \G{4}-\Gv{4} + \La{4} \Gt (\G{4}-\Gv{4}) \\
  & \quad + (\G{4}-\Gv{4}) \Gtv \Lv{4} \\
  & \quad +\La{4}(\Gt-\Gtv) \Lv{4}  \\
  & \quad + \La{4} \Gt(\G{4}-\Gv{4}) \Gtv \Lv{4}
\label{BS4-finite}
\end{split}
\end{equation}
Since the $\G{4}$-functions are 2PI when cutting the corresponding diagrams
such that the two external point pairs become separated, the subtractions
cause the explicite loops in the final expression to be finite such that
only the vacuum quantities are to be renormalized.

This vacuum renormalization is obtained with the same techniques as
explained in {\I}. In the case of a spontaneously broken symmetry only the
renormalization description has to be changed, since the on-shell scheme
described in {\I} would lead to artificial infrared singularities.  Rather
here a ``mass-independent renormalization scheme'' as the MS or
$\overline{\text{MS}}$ scheme in dimensionally regularized perturbation
theory is required. As generally shown in {\I}, for self-consistent schemes
BPHZ-like renormalizations are more convenient to use, especially in
numerical simulations as presented in {\II}. Thus we introduce a mass
renormalization scale $\tilde{\mu}$, i.e., \emph{for the counter terms} we
set the mass parameter $m^{2}$ to $\tilde{\mu}^{2}$. Since the UV
divergences are ruled by the asymptotic behavior of the Green's functions
at large loop momenta, the same counter terms also render the integrals
finite for the spontaneously broken phase where $m^2=-\tilde{m}^{2}<0$.

Then we can apply the same renormalization conditions as in {\I} for the
symmetric phase now however taken at $m^2=\tilde{\mu}^{2}$:
\begin{equation}
\label{5n.17}
\begin{split}
  \Sigma^{(\text{vac})}(p^{2}=0;m^{2} &= \tilde{\mu}^{2}) = 0, \\
  \partial_{p^{2}}
  \Sigma^{(\text{vac})}(p^{2};m^{2} &= \tilde{\mu}^{2})|_{p^{2}=0}=0, \\
  \Lv{4}(s=t=u=0;m^2=\tilde{\mu}^{2}) &=
  \Gv{4}(s=t=u=0;m^{2}=\tilde{\mu}^{2}) = \frac{\lambda}{2}.
\end{split} 
\end{equation}
Herein $s$, $t$ and $u$ are the usual Mandelstam variables for
two-particle scattering kinematics. This procedure shows that those
counter-terms necessary to render the self-consistent Dyson-equation
of motion finite also enter here for the BS-equation.  Especially the
crossing symmetry of counterterms is also recovered consistently with
the chosen approximation for $\Phi$.

Given the renormalized $\Lambda^{(4;\text{vac})}$, Eq.
(\ref{BS4-finite}) constitutes a regular integral equation to
determine the finite $T$-dependent function $\Lambda^{(4)}$.  The
steps towards $\Lambda^{(3)}$ and finally $\Sigma^{\text{BS}}$ only
involve to close a further loop which is rendered UV-finite
considering the respective differences
\begin{equation}
\label{BS3-finite}
\begin{split}
  \La{3}-\Lv{3}=&\G{3}-\Gv{3}+
  \underbrace{\La{4} \tilde{G}^{(2)} \G{3}-\Lv{4}
  \Gtv \Gv{3}}_{\text{UV-finite}} \\
  \Sigma_{\text{BS}}-\Sigma_{\text{BS}}^{\text{vac}}=&
  \underbrace{\G{3} \tilde{G}^{(2)}
  \La{3} - \Gv{3} G^{(2;\text{vac})}\Lv{3}}_{\text{UV-finite}}.
\end{split}
\end{equation}
Here it is important to notice that in purely scalar theories without
derivative couplings three-point vertices are only logarithmically
divergent. Further due to field-reflection symmetry of a theory with no
generic three-point interaction term in the classical action no such
counter term is necessary to render the three-point functions
$\G{3}$ and $\La{3}$ finite; only counter terms for four-point
functions appear in the renormalization scheme
cf. (\ref{BS3-finite}).

This completes the proof that also for the non-perturbative higher
order vertex functions resulting from $\tilde{\Gamma}$ the
renormalization parts can entirely be defined at the vacuum level.

\subsection{Comments}

The main result of our symmetry analysis can be summarized as follows:
Truncated expansions of the 2PI \emph{functional} $\Gamma[\vec{\varphi},G]$
with respect to parameters, like $\hbar$ or $1/N$, consistent with the
linearly realized symmetries yield approximate functionals which are
symmetric, when mean fields and Green's functions are transformed as
O($N$)-vectors and 2$^{\text{nd}}$-rank O($N$)-tensors, respectively.
However, the solutions of the dynamical equations of motion may imply a
resummation to arbitrary orders in the considered expansion
parameter\footnote{Exceptions are truncations right at zero order as all
  zero-order terms are self-replicative, see the $1/N$-expansion discussed
  in sect. \ref{subsec-1/N}}. Then this resummation is incomplete and
crossing and O($N$)-symmetry may be violated at orders beyond those
explicitely included in the chosen approximant of $\Phi$.  This includes a
violation of the $\beta$-function of the running coupling constant at the
very same order (see also \cite{bp01} and {\II}).

The techniques developed in {\I} can be used to renormalize the
self-consistent self-energy and the 2PI $\Gamma$-functional with
temperature-independent counter terms. As usual in the here considered case
of a spontaneously broken symmetry the $\Gamma$-functional can be rendered
finite at a mass scale $\tilde{\mu}^2>0$, i.e., in the symmetric
Wigner-Weyl phase. Thus no particular problems with massless degrees of
freedom appearing in the Nambu-Goldstone mode arise, while symmetries are
violated at the self-consistent level.

The symmetry is appropriately repaired through the non-perturbative
approximation (\ref{4.2}) for the 1PI effective action functional. It was
shown that all symmetries are recovered in the sense that the approximate
proper vertex functions defined from (\ref{4.3}) in the usual sense, i.e.,
by
\begin{equation}
\tilde{\Gamma}_{1,\ldots,n} = \ii \frac{\delta^{n}
  \tilde{\Gamma}[\varphi]}{\delta \varphi_{1} \cdots \delta \varphi_{n}}
\end{equation}
fulfill the usual WTIs of the proper vertex-functions and can be
renormalized with temperature independent counter terms by the techniques
developed in {\I} and Sect. \ref{sect-ren-bs}.

Based on the symmetry violating intermediate Dyson resummation, however, it
is expected that certain relics of that \emph{internal} stage are still
present also in these symmetric proper vertex functions\footnote{This
    is essentially true for all approximations other then self-consistent
    ones, e.g., in perturbation theory the internal structure is given by
    free particle properties!}. Since the propagators, which define the
kernels of Bethe-Salpeter equations, are given by the symmetry violating
self-consistent scheme, already the threshold structure is expected to
deviate from the correct behavior, since the Nambu-Goldstone modes falsely
appear with finite masses for these \emph{internal} lines.

One could think to include the external propagator into the self-consistent
scheme, i.e., to consider all quantities as a function of $G_{\text{ext}}$
rather than the Dyson $G$. This could indeed further improve the
approximation. However, since implicitly this defines a new
$\Phi$-derivable scheme, with a $\Phi$ defined by the 2PI diagrams leading
to $\Sigma_{\text{ext}}$, it has the consequence that the symmetries may
again be violated, although at a much more ``remote'' level.

\section{Approximations of the $\Phi$-functional}
\label{sect-on-funct}

In this section we first give a functional derivation of the
$\Phi$-functional with the techniques developed in Sect.
\ref{sect-baym-funct} up to order $\hbar^{2}$ to exemplify the reduction to
2PI diagrams and the symmetry properties of the approximations for the
functional. As an application we give a diagrammatic derivation for the
functional up to order $\lambda^{2}$.

We write the Lagrangian of the O($N$)-model in the following form:
\begin{equation}
\label{5n.1}
\Lag=\frac{1}{2} (\partial_{\mu} \vec{\phi}) (\partial^{\mu} \vec{\phi})
+\frac{\tilde{m}^{2}}{2} \vec{\phi}^{2} - \frac{\lambda}{8}
(\vec{\phi}^{2})^2.
\end{equation}
It is immediately clear that in this case for both, the loop expansion
and the coupling constant expansion, the corresponding approximations
for the $\Phi$-functional are symmetric under O($N$)-transformations,
where $\varphi$ and $G$ transform as a vector and a tensor of
2$^{\text{nd}}$ rank, respectively.

\subsection{The $\Phi$-functional up to order $\hbar^{2}$}

According to our discussion in Appendix \ref{sect-baym-funct} we have to
calculate the functional
\begin{equation}
\label{5n.2}
\begin{split}
  Z_{1}[J'] = & N \exp \left ( \frac{\ii}{\hbar} \tilde{S}[\varphi_{0}',J']
  \right ) \int \D \Phi'' \exp \Bigg \{ \frac{\ii}{2}
  \funcint{(\mathscr{G}^{-1})_{1j,2k} (\phi'')_{1}^{j} (\phi'')_{2}^{k}}{12} \\
  & - \frac{\ii \lambda}{24} \delta_{jklm} \funcint{\left (\hbar
      (\phi'')_{1}^{j} (\phi'')_{1}^{k} (\phi'')_{1}^{l} (\phi'')_{1}^{m} +
      4 \sqrt{\hbar} (\phi'')_{1}^{j} (\phi'')_{1}^{k} (\phi'')_{1}^{l}
      \varphi_{1}^{m} \right)}{1} \Bigg \},\\
  & \quad\text{where}\quad
  \delta_{jklm} =  \delta_{jk} \delta_{lm} +  \delta_{jl} \delta_{km}
  + \delta_{jm} \delta_{kl}  
\end{split}
\end{equation}
and where we have explicitly reintroduced the internal field indices. The
$\hbar$-expansion can now be generated following the standard technique for
the usual perturbative calculation (see, e.g., \cite{itz80}). All path
integrals of the $\hbar$ expansion can be found be taking functional
derivatives of the Gaussian path integral
\begin{equation}
\label{5n.3}
\begin{split}
  Z_{10}[J',K] &= N \int \D \Phi'' \exp \left (
    \funcint{(\mathscr{G}^{-1})_{1j,2k}
      (\phi'')_{1}^{j} (\phi'')_{2}^{k}}{12} \right) \\
  &= \exp \left [-\frac{1}{2} \Tr \ln(\mathscr{G}^{-1}/M^{2}) -\frac{\ii}{2}
    \funcint{\mathscr{G}_{12}^{jk} K_{1j} K_{2k}}{12} \right ]
\end{split}
\end{equation}
with respect to the new independent local auxiliary source $K$ which has to
be set to $0$ at the end of the calculation. We have to substitute only
\begin{equation}
\label{5n.4}
(\phi'')_{1}^{j} \rightarrow \frac{1}{\ii} \frac{\delta}{\delta K_{1j}}
\end{equation}
in the polynomial expression for the expansion of $\exp (S_{I}[\sqrt{\hbar}
\phi'',\varphi] )$.

In our case this leads to
\begin{equation}
\label{5n.5}
\begin{split}
W_{1}[J'] &= -\ii \hbar \ln [Z_{10}[J',K=0](1 + \hbar z_{1}^{[1]} + \hbar^{2}
z_{1}^{[2]})] + O(\hbar^{3}) \\
 &= \tilde{S}[\varphi_{0}',J']+ \frac{\ii \hbar}{2} \Tr \ln
 (\mathscr{G}^{-1} M^{2}) \\ & \quad \quad + \hbar^{2} \left ( \frac{\lambda}{8}
   \delta_{jklm} \funcint{\mathscr{G}_{11}^{jk} \mathscr{G}_{11}^{lm}}{1} +
   \frac{\lambda^{2}}{8} \delta_{jklm} \delta_{j'k'l'm'} \funcint{
\varphi_{1}^{j} \mathscr{G}_{11}^{kl} \mathscr{G}_{12}^{mm'}
\mathscr{G}_{22}^{k'l'} \varphi_{2}^{j'}}{12} \} \right).
\end{split}
\end{equation}
Now we need to substitute $\varphi'=0$ instead of $\varphi_{0}'$ in this
expression. To that end we have to find the expression for
\begin{equation}
\label{5n.6}
\varphi'=\funcd{W_{1}[J']}{J'}
\end{equation}
up to order $\hbar$. Since $\varphi_{0}$ is the stationary point of
$\tilde{S}[\phi',J']$ at fixed $J'$ we have
\begin{equation}
\label{5n.7}
\funcd{\tilde{S}[\varphi_{0}',J']}{J'}=\varphi_{0}'.
\end{equation}
Further from the equation of motion for $\varphi_{0}'$ we have
\begin{equation}
\label{5n.8}
\funcd{(\varphi_{0}')_{1}^{j}}{(J')_{2k}} = \left  [ \left(
      \funcd{J'}{\varphi_{0}'} \right)^{-1} \right ]_{12}^{jk} =
      \mathscr{G}_{12}^{jk}. 
\end{equation}
From this we get
\begin{equation}
\label{5n.9}
\frac{\ii \hbar}{2} \frac{\delta}{\delta (J')_{1}^{j}} \Tr \ln
(\mathscr{G}^{-1}/M^{2}) = \frac{\ii \hbar}{2} \lambda \delta_{klmn}
\funcint{G_{1'1'}^{kl} \varphi_{1'}^{m} G_{1'1}^{nj}}{1'} + O(\hbar^2)
\end{equation}
and thus
\begin{equation}
\begin{split}
  \frac{\ii \hbar}{2} \Tr \ln(\mathscr{G}^{-1}/M^2) = & \frac{\ii \hbar}{2}
  \Tr \ln(G^{-1} M^2) \\ & -\frac{\hbar^{2}}{4} \lambda^{2} \delta_{jklm}
  \delta_{j'k'l'm'} \funcint{\varphi_{1}^{j} G_{11}^{kl} G_{12}^{mm'}
    G_{22}^{l'k'} \varphi_{2}^{j'}}{12} + O(\hbar^{3})
\end{split}
\end{equation}
and
\begin{equation}
\label{5n.10}
\tilde{S}[\varphi_{0}']=\frac{\hbar^{2}}{8} \lambda^{2} \delta_{jklm}
\delta_{j'k'l'm'} \funcint{\varphi_{1}^{j} G_{11}^{kl} G_{12}^{mm'}
  G_{22}^{l'k'} \varphi_{2}^{j'}}{12} + O(\hbar^{3}) 
\end{equation}
Gathering all terms finally yields
\begin{equation}
\label{5n.11}
\begin{split}
  \Gamma[\varphi,G] &= S[\varphi] - \frac{\ii \hbar}{2}
  \funcint{\mathscr{D}_{12}^{-1} (G_{12}-\mathscr{D}_{12})}{12} +
  W_{1}[J_{0}'] \\
  &= S[\varphi] + \frac{\ii \hbar}{2} \Tr \ln(G_{12}^{-1}/M^{2}) +
  \frac{\hbar^{2} \lambda}{8} \delta_{jklm} \funcint{G_{11}^{jk}
    G_{11}^{lm}}{1} \\ & \quad \quad + \frac{\hbar^{2} \lambda^{2}}{12}
  \delta_{jklm} \delta_{j'k'l'm'} \funcint{\varphi_{1}^{j} G_{12}^{kk'}
    G_{12}^{ll'} G_{12}^{mm'} \varphi_{2}^{j'}}{12} + O(\hbar^{3}).
\end{split}
\end{equation}
This is indeed invariant under global O($N$) operations if the mean
fields $\varphi$ and $G$ are transformed as tensors of 1$^{\text{st}}$ and
2$^{\text{nd}}$ rank respectively.

\subsection{The $\Phi$-functional to order $\lambda^{2}$}

Now we like to give an example for the application of the diagram rules for
the $\Phi$-functional and the equations of motion for the self-consistent
field and self-energy.

From (\ref{2.10}) we see that the $\Phi$-functional is given by diagrams
which obey the Feynman rules with the point vertices of a field theory with
an interaction Lagrangian given by $S_{I}[\phi,\varphi]$. The Lines,
connecting these point vertices denote full propagators $\ii G$ rather than
perturbative propagators. As argued in the paragraph after equation
(\ref{2n.17}) one has to keep only two-particle irreducible closed diagrams
with at least two loops for $\ii \Phi$.

For the O($N$)-model this leads to the following diagram expression
for $\ii \Phi$ up to order $\lambda^{2}$ in the coupling-constant expansion
scheme
\begin{equation}
\label{5n.12}
\ii \Phi[\varphi,G] = \parbox{5.3cm}{\includegraphics{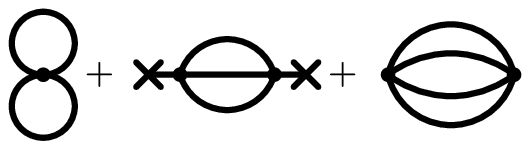}}.
\end{equation}
The corresponding analytic expression reads
\begin{equation}
\label{5n.13}
\begin{split}
  \Phi[\varphi,G] = & \frac{\hbar^{2} \lambda}{8} \delta_{jklm}
  \funcint{G_{11}^{jk} G_{11}^{lm}}{1} \\
  & + \frac{\hbar^{2} \lambda^{2}}{12} \delta_{jklm} \delta_{j'k'l'm'}
  \funcint{\varphi_{1}^{j} G_{12}^{kk'} G_{12}^{ll'} G_{12}^{mm'}
    \varphi_{2}^{j'}}{12} \\
  & + \frac{\ii \hbar^{3} \lambda^{2}}{48} \delta_{jklm} \delta_{j'k'l'm'}
  \funcint{G_{12}^{jj'} G_{12}^{kk'} G_{12}^{ll'} G_{12}^{mm'}}{12} +
  O(\lambda^{3}).
\end{split}
\end{equation}
The equations of motion for the approximation are given by
(\ref{2n.16}-\ref{2n.17}). For the mean field we find
\begin{equation}
\label{5n.14}
\begin{split}
  \ii(-\Box+\tilde{m}^{2}) \varphi_{1}^{j} - \frac{\ii \lambda}{6}
  \delta_{jklm} \varphi_{1}^{k} \varphi_{1}^{l} \varphi_{1}^{m} 
  + \frac{\hbar \lambda}{2} \delta_{j'k'jm} \varphi_{1}^{m} G_{11}^{j'k'} \\ +
  \frac{\ii \hbar^{2} \lambda^{2}}{6} \delta_{jklm} \delta_{j'k'l'm'}
  \funcint{G_{12}^{kk'} G_{12}^{ll'} G_{12}^{mm'} \varphi_{2}^{j'}}{2} &
  \stackrel{!}{=}0.
\end{split}
\end{equation}
In terms of diagrams this equation of motion reads
\begin{equation}
\label{5n.15}
-\ii (-\Box+\tilde{m}^{2}) \varphi_{1} = \raisebox{0.5mm}{
 \parbox{4.4cm}{\includegraphics{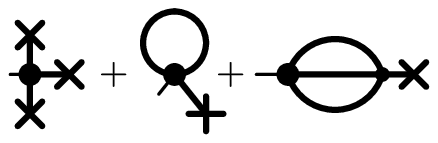}}}.
\end{equation}
The equation for the self-consistent self-energy follows from (\ref{3n.8})
\begin{equation}
\label{5n.16}
\begin{split}
  -\ii \Sigma_{1j,2k} &= \frac{\ii \hbar \lambda}{2} \delta_{jklm}
  G_{11}^{lm} \delta_{12}^{(d)} + \frac{\ii \hbar \lambda^{2}}{2}
  \delta_{jpqr} \delta_{kp'q'r'} \varphi_{1}^{p} G_{12}^{rr'} G_{12}^{qq'}
  \varphi_{2}^{p'} \\
  &\quad - \frac{\hbar^{2} \lambda^{2}}{6} \delta_{jpqr} \delta_{kp'q'r'}
  G_{12}^{pp'} G_{12}^{qq'} G_{12}^{rr'} \\
&= \raisebox{1.6mm}{\parbox{5.0cm}{\includegraphics{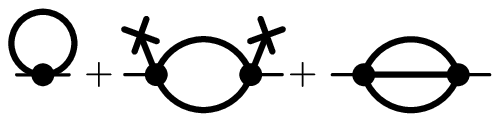}}}.
\end{split}
\end{equation}
According to (\ref{4.14}-\ref{4.16}) the diagrammatical elements for
the calculation of the external self-energy 
are given by
\begin{alignat}{4}
  \frac{\partial \ii \Phi[\varphi,G]}{\delta \varphi_{1} \delta
    \varphi_{2}} &=&
  \raisebox{0.14mm}{\parbox{1.75cm}{\includegraphics{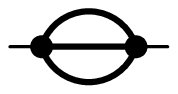}}}
  \;,& & & &\\
  \ii \Gamma^{(3)} &=&
  \raisebox{-1.5mm}{\parbox{0.9cm}{\includegraphics{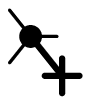}}}
  \hspace{4mm} &+ \hspace{5mm}&
  \raisebox{1.5mm}{\parbox{0.8cm}{\includegraphics{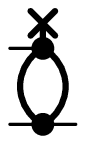}}}, \hspace{8mm} & &\\
  \ii \Gamma^{(4)} &=& \parbox{0.6cm}{\includegraphics{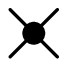}}
  \hspace{6.5mm}&+ \hspace{5mm} & \parbox{0.9cm}{\includegraphics{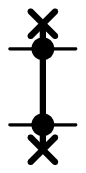}}
    \hspace{8mm} & \hspace{-5mm}+ \hspace{4mm}& 
  \parbox{1.0cm}{\includegraphics{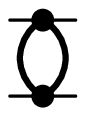}}.
\end{alignat}

\section{A simple example: The Hartree (plus exchange) approximation}
\label{sect-example}

To illustrate the above formal considerations in this section we
numerically solve the equations of motion for both the self-consistent and
the symmetric effective self-energies for the approximation keeping only
terms to linear order in the explicitely appearing coupling $\lambda$ in
all above equations. This defines the Hartree approximation including the
corresponding bosonic exchange terms\footnote{The latter are of
  next-to-leading order in the semi-classical $1/N$ expansion.}.
Especially $\Phi$, c.f.  (\ref{5n.13}), becomes
\begin{equation}
\label{6n.1}
\Phi[\varphi,G]=\frac{\lambda}{8} \delta_{jklm}
  \funcint{G_{11}^{jk} G_{11}^{lm}}{1},
\end{equation}
where we have set $\hbar=1$. 

\subsection{The vacuum case}

In the following we can restrict ourselves to time ordered functions, so
that in this section all propagators stand for $\{--\}$-propagators and all
vertices for $\{-\}$-vertices. 

First we have to find and solve the self-consistent equations of
motion (\ref{3n.7}) for the approximation given by the $\Phi$
functional (\ref{6n.1}). For a given mean field $\varphi_{1}^{j}$ we
find:
\begin{equation}
\label{6.1}
(\mathscr{D}^{-1})_{1j,2k}=\left [ (-\Box_{1}+\tilde{m}^{2}) \delta_{jk} -
  \frac{\lambda}{2} (2 \varphi_{1j} \varphi_{1k}+\vec{\varphi}_{1}^{2}
  \delta_{jk})\right] \delta^{(d)}(x_{1}-x_{2}).
\end{equation}
Since the vacuum is homogeneous in space and time the mean field is
constant and pointing in one arbitrarily chosen direction. Therefore it is
convenient to express all vector and tensor quantities in terms of their
components parallel and perpendicular to $\vec{\varphi}$. The perpendicular
modes are those of the Goldstone bosons (e.g., the pions in the linear
sigma model).  The corresponding projectors are
\begin{equation}
\label{6.2}
\begin{split}
  P_{\perp}^{jk} &= \delta^{jk}-\frac{\varphi^{j}
    \varphi^{k}}{\vec{\varphi}^{2}}, \\
  P_{\parallel}^{jk} &=\frac{\varphi^{j} \varphi^{k}}{\vec{\varphi}^{2}}.
\end{split}
\end{equation}

In momentum representation the self-energy components derived from the
approximation to $\Phi$ given by (\ref{6n.1}) become
\begin{equation}
\label{6.3}
\begin{split}
\Sigma_{\perp} &= \frac{\ii \lambda}{2} \mu^{2 \epsilon} \feynint{l} [(N+1)
G_{\perp}(l) + G_{\parallel}(l)], \\
\Sigma_{\parallel} &= \frac{\ii \lambda}{2} \mu^{2 \epsilon} \feynint{l} [(N-1)
G_{\perp}(l) + 3 G_{\parallel}(l)].
\end{split}
\end{equation}
Correspondingly the Dyson equations (\ref{3n.7-approx}) decouple with
\begin{equation}
\label{6.4}
\begin{split}
& G_{\perp}(p) = \frac{1}{p^{2}-M_{\perp}^{2}+\ii \eta}, \\
& G_{\parallel}(p) = \frac{1}{p^{2}-M_{\parallel}^{2}+\ii \eta} \\
\text{with } & M_{\perp}^{2}=\frac{\lambda}{2} \vec{\varphi}^{2} -
\tilde{m}^{2} + \Sigma_{\perp}, \\
& M_{\parallel}^{2}=\frac{3 \lambda}{2} \vec{\varphi}^{2} -
\tilde{m}^{2} + \Sigma_{\parallel}.
\end{split}
\end{equation}
With these definitions the equation of motion for the mean field reads
\begin{equation}
\label{6.5}
\varphi_{l}(M_{\parallel}^{2}-\lambda \vec{\varphi}^{2})=0,
\end{equation}
which shows that either the mean field or the bracket vanishes. In
the first case, the symmetric Wigner-Weyl mode, in the second the
Nambu-Goldstone mode is realized. In our case of negative $-\tilde{m}^{2}$
the spontaneously broken phase is realized in the vacuum.

In the following we shall use the convention and renormalization
prescription according to (\ref{5n.17}). In this renormalization
scheme the self-consistent gap equations (\ref{6.3}-\ref{6.4})
become
\begin{equation}
\label{6.6}
\begin{split}
  M_{\perp}^{2} = \TP_{\perp}(M_{\perp},M_{\parallel}):=\frac{\lambda}{2}
  \vec{\varphi}^{2} - \tilde{m}^{2} + \frac{\lambda}{32 \pi^{2}} \Bigg [ &
  (N+2) \tilde{\mu}^{2} -(N+1)M_{\perp}^{2} - M_{\parallel}^{2} \\
  & +M_{\perp}^{2} (N+1) \ln \left (\frac{M_{\perp}^{2}}{\tilde{\mu}^{2}}
  \right ) + M_{\parallel}^{2}\ln \left
    (\frac{M_{\parallel}^{2}}{\tilde{\mu}^{2}}
  \right ) \Bigg ] \\
  M_{\parallel}^{2} = \TP_{\parallel}(M_{\perp},M_{\parallel}):= \frac{3
    \lambda}{2} \vec{\varphi}^{2} - \tilde{m}^{2}+\frac{\lambda}{32
    \pi^{2}} \Bigg [ & (N+2) \tilde{\mu}^{2} -(N-1)M_{\perp}^{2} -3
  M_{\parallel}^{2} \\ & + M_{\perp}^{2} (N-1) \ln \left
    (\frac{M_{\perp}^{2}}{\tilde{\mu}^{2}} \right ) + 3
  M_{\parallel}^{2}\ln \left (\frac{M_{\parallel}^{2}}{\tilde{\mu}^{2}}
  \right ) \Bigg ]
\end{split}
\end{equation}
The equations show clearly that the physical results are independent of the
choice of the mass renormalization scale $\tilde{\mu}$ since a different
choice of $\tilde{\mu}$ can be compensated by a (finite) renormalization of
the coupling $\lambda$ and the mass parameter $\tilde{m}$. For more general
$\Phi$-derivable approximations, which contain real two-point contributions
to the self-energy, also a wave function renormalization is needed
(see \cite{vHK2001-Ren-I}). The introduction of the mass renormalization
scale $\tilde{\mu}$ is necessary only because we need a ``\emph{mass
  independent renormalization scheme}'' \cite{hooft73,bard78} which uses
the fact that a theory with symmetries can be renormalized with counter
terms of its symmetric realization in the Wigner-Weyl mode also for the
spontaneously broken Nambu-Goldstone phase. In addition mass independent
renormalization schemes have the advantage to avoid ``renormalization
induced'' infrared divergences which would appear with the massless modes
within an on-shell renormalization scheme. It is also clear that for
special choices of $\tilde{\mu}$ our class of renormalization schemes
contains the minimal subtraction scheme (MS) and the modified minimal
subtraction scheme ($\overline{\text{MS}}$).

From (\ref{6.6}) it becomes immediately clear that by the self-consistent
$\Phi$-derivable approximation the symmetry is explicitly broken in the
Nambu-Goldstone mode of the theory, i.e., for $m^{2}=-\tilde{m}^{2}<0$
which provides $\varphi^{2}>0$: The transverse modes should be massless
according to Goldstone's theorem which does not hold true according to
(\ref{6.6}).

As we have seen in section \ref{sect-eff-act} we can define an
approximation to the self-energy respecting the underlying symmetry by
(\ref{4.8}). For that purpose we need to solve the equation of motion
(\ref{4.13}). From (\ref{4.14}) we obtain the kernels, here given in the
momentum representation
\begin{equation}
\label{6.7}
\begin{split}
  \Gamma_{jk;l,m}^{(4)} &= -\frac{\lambda}{2} (\delta_{jk} \delta_{lm} +
  \delta_{jl} \delta_{km} + \delta_{jm} \delta_{kl}) :=
  -\frac{\lambda}{2} \delta_{jklm}, \\
  \Gamma_{jk;l}^{(3)}&= -\lambda (\varphi_{k} \delta_{jl} + \varphi_{j}
  \delta_{kl} + \varphi_{l} \delta_{jk})=-\lambda \delta_{jklm}
  \varphi_{m}.
\end{split}
\end{equation}
Since due to translation invariance $\varphi$ is independent of the
four-momentum both $\Gamma^{(4)}$ and $\Gamma^{(3)}$ are constant and
effectively only one-point functions. From this it is clear that
$\Lambda^{(3)}$ is effectively only a two-point function and thus depends
only on the momentum attached to its third argument.  This becomes also
immediately clear from the diagrams (\ref{4.13b}) and (\ref{4.16}). In our
special case the kernels $\Gamma^{(3)}$ and $\Gamma^{(4)}$ are point
vertices
\begin{equation}
\label{6.8}
\ii
\Gamma^{(3)}=\raisebox{-1.6mm}{\parbox{0.9cm}{\includegraphics{IPhiTad.eps}}}
\text{ and } \; \ii \Gamma^{(4)} = \parbox{0.9cm}{\includegraphics{KPhieight.eps}}
\end{equation}
and thus from the iterative solution of the equation of motion
(\ref{4.13}) and (\ref{4.15}) for $\Lambda^{(3)}$ we conclude that the
external self-energy associated with the Hartree approximation is
given by the \emph{RPA bubble resummation}
\begin{equation}
\label{6.13b}
-\ii
 \Sigma_{\text{ext}}=\raisebox{1.6mm}{\parbox{7.6cm}{\includegraphics{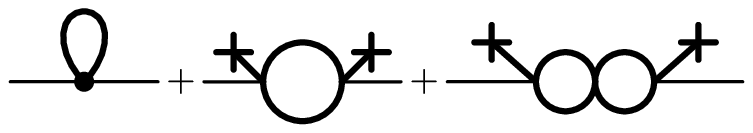}}}
 + \cdots
\end{equation}
To find the analytic expressions for this diagrammatic equation we
specialize (\ref{4.13}) with the kernels (\ref{6.7}). This leads to the
equation
\begin{equation}
\label{6.9}
\Lambda^{(3)}_{jkl}(k) = -\lambda \delta_{jklm} \varphi^{m} + \frac{\ii
  \lambda}{2} \delta_{jkl'm'} \Lambda_{l''m'';l}^{(3)}(k) \feynint{l}
  G^{l'l''}(l) G^{m'm''}(l-k).
\end{equation}
According to (\ref{5n.17}) for renormalization we have to subtract
the expression where in (\ref{6.9}) both propagators are set to free
propagators with the mass parameter set to the mass renormalization scale
$\tilde{\mu}$ and $k=0$.

From the symmetry properties of $\Lambda^{(3)}$ it follows that it must be
of the form
\begin{equation}
\label{6.10}
\begin{split}
  \Lambda^{(3)}_{jk;l}(k) =&\varphi^{l} \left (\Lambda_{1}^{(3)}(k)
    P_{\perp}^{jk} + \Lambda_{2}^{(3)}(k) P_{\parallel}^{jk} \right ) \\
  &+ \Lambda_{3}^{(3)}(k) \left(\varphi^{j} P_{\perp}^{kl} + \varphi^{k}
    P_{\perp}^{jl} \right ) \\
  &+ \Lambda_{4}^{(3)}(k) \left(\varphi^{j} P_{\parallel}^{kl} + \varphi^{k}
    P_{\parallel}^{jl} \right ).
\end{split}
\end{equation}
Indeed after some algebraic manipulations with the propagators (\ref{6.4})
we obtain the algebraic linear equations of motion for the four independent
components of $\Lambda^{3}$ as
\begin{equation}
\label{6.11}
\begin{split}
  \Lambda_{1}^{(3)}(k) &= -\lambda + \frac{\lambda}{2} \left [(N+1) L_{\perp,
      \perp}(k) \Lambda_{1}^{(3)}(k) + (\Lambda_{2}^{(3)}(k)+2
    \Lambda_{4}^{(3)}(k)) L_{\parallel, \parallel}(k) \right], \\
  \Lambda_{2}^{(3)}(k) &= -\lambda + \frac{\lambda}{2} \left [(N-1) L_{\perp,
      \perp}(k) \Lambda_{1}^{(3)}(k) + (3 \Lambda_{2}^{(3)}(k)+2
    \Lambda_{4}^{(3)}(k)) L_{\parallel, \parallel}(k) \right], \\
  \Lambda_{3}^{(3)}(k) &= -\lambda + \lambda L_{\parallel, \perp}(k)
  \Lambda_{3}^{(3)}(k), \\
  \Lambda_{4}^{(4)}(k) &= -\lambda + \lambda L_{\parallel, \parallel}(k)
  \Lambda_{4}^{(3)}(k).
\end{split}
\end{equation}
Here we used the abbreviation
\begin{equation}
\label{6.12}
\begin{split}
  & L_{\alpha, \beta}(k) = L_{\beta,\alpha}(k) = \ii \fint{l} \left \{
    G_{\alpha}(l) G_{\beta}(l-k) - [D(l)]^{2} \right \} \\ \text{with } &
  \alpha,\beta \in \{\perp,\parallel \} \text{ and }
  D(l)=\frac{1}{l^{2}-\tilde{\mu}^{2} + \ii \eta}.
\end{split}
\end{equation}
According to (\ref{4.15}) and (\ref{6.9}) the external self-energy is given
by
\begin{equation}
\label{6.13}
\begin{split}
  (\Sigma_{\text{ext}})_{jk}(p) &= \Sigma_{jk}(p)-\varphi^{l}
  (\Lambda_{lj;k}^{(3)}(p)-\Gamma_{ljk}^{(3)})  \\
  &= \Sigma_{jk} - \vec{\varphi}^{2} \{ [3 \lambda +
  \Lambda_{2}^{(3)}(p)+2\Lambda_{4}^{(3)}(p)] P_{\parallel jk}+
  [\lambda+\Lambda_{3}^{(3)}(p)] P_{\perp jk} \} .
\end{split}
\end{equation}
In order to prove Goldstone's theorem we have to set $p=0$ and use the
equations of motion (\ref{6.5}) and (\ref{6.6}). The crucial property is
\begin{equation}
\label{6.14}
M_{\perp}^{2} =
\Sigma_{\perp}-\Sigma_{\parallel} =
-\lambda[\TP_{1}(M_{\parallel}^2)-\TP_{1}(M_{\perp}^{2}) 
- (M_{\parallel}^{2}-M_{\perp}^{2}) \TP_{2}(\tilde{\mu}^2)],
\end{equation}
where the functions $\TP_{1}$ and $\TP_{2}$ are given in appendix \ref{app-b}
cf. (\ref{b.1}) and (\ref{b.2}). Using (\ref{b.5}) together with
\begin{equation}
\label{6.15}
M_{\text{ext}\perp}^{2}=\frac{\lambda}{2} \vec{\varphi}^{2}-\tilde{m}^{2} +
\Sigma_{\text{ext} \perp}(0)
\end{equation}
and (\ref{6.13}) we find indeed $M_{\text{ext}\perp}^{2}=0$ as it should be
according to Goldstone's theorem: The $(N-1)$ degrees of freedom
perpendicular to the mean field are the Goldstone modes corresponding to
the spontaneous breaking of the O($N$) to the O($N-1$) symmetry group of
the mean field.

Finally we remark that the Hartree self-energy for the case of the
Wigner-Weyl phase of the theory, where $\vec{\varphi}=0$, is identical with
the external self-energy and thus fulfills also the WTI for the
self-energy, which indeed in this case is trivially fulfilled.

\begin{figure}
\begin{minipage}{0.45\textwidth}
\centerline{\includegraphics[height=70mm]{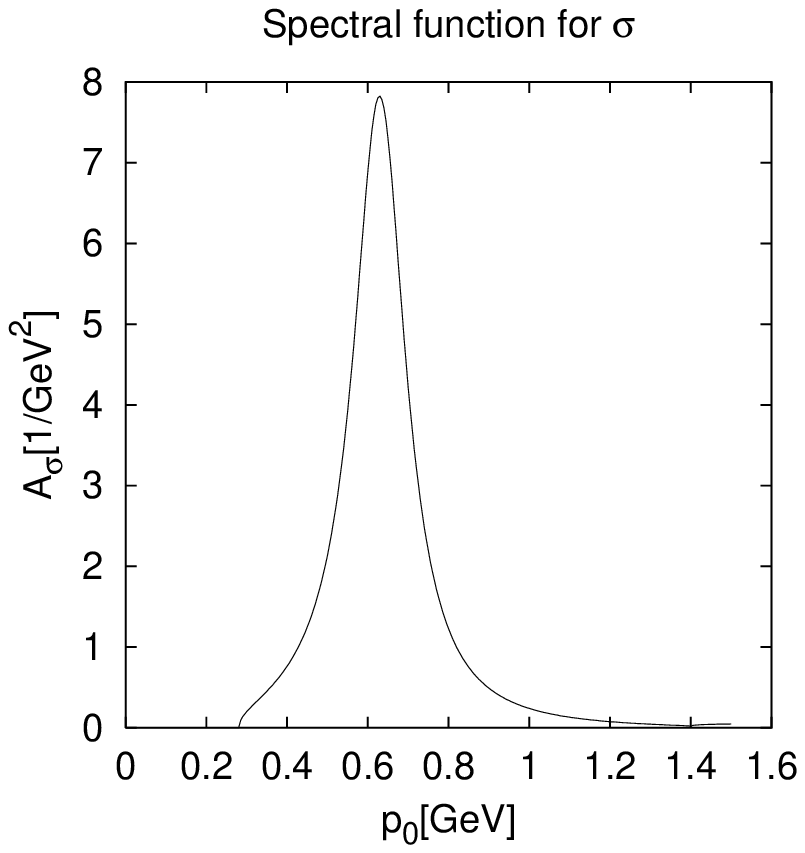}}
\end{minipage}
\begin{minipage}{0.45\textwidth}
\centerline{\includegraphics[height=70mm]{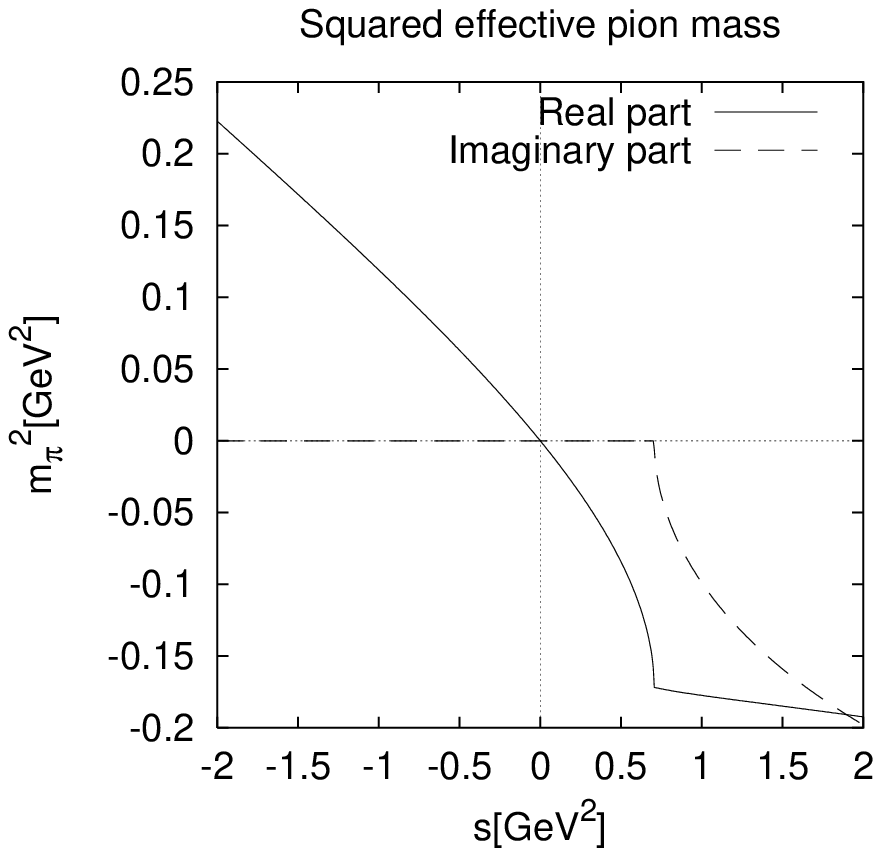}}
\end{minipage}
\caption{The ``$\sigma$'' spectral function (left) and the external effective
  mass of the ``pions'' (right).}
\label{fig.1}
\end{figure}

In Fig. \ref{fig.1} we plot an example for the above considerations: We
roughly fit the parameters of the self-consistent self-energy such that the
properties for pions are satisfied, namely $M_{\perp}=140 \text{MeV}$ and
$M_{\parallel}=600 \text{MeV}$, $\varphi=f_{\pi}=93 \text{MeV}$, and
$n=4$. The equations of motion (\ref{6.6}) were solved for $\tilde{m}^{2}$
and $\tilde{\mu}^{2}$. The renormalization conditions (\ref{5n.17}) were
used.
The plot of the pion mass clearly shows that Goldstone's theorem is
recovered, since $(M_{\perp}^{\text{(ext)}})^2(s=0)=0$ although the internal
Green's functions do not fulfill the Ward identities. This violation of the
symmetry properties by the self-consistent self-energies is clearly seen in
the spectral function for the ``$\sigma$-meson'', since its threshold is at
$\sqrt{s}=0 = 2 M_{\perp}=280 \text{MeV}$ and not at $\sqrt{s}=0$ as it
should be for massless pions in the chiral limit.

Like the here constructed \emph{external} self-energy where due to the
finite mean field $\varphi$ the RPA-bubbles contributed to the self-energy
expectation value $\varphi(x)\varphi(y)\left<\phi^2(x)\phi^2(y)\right>$
also other symmetry preserving two-point functions are given by the very
same RPA terms, and thus result from the linear response of the system due
to fluctuations around the Hartree solution. A prominent example is the
correlator $\left<j^{\mu}(x)j^{\nu}(y)\right>$ of the Noether current
(\ref{3.7}), which then is conserved.

\subsection{The finite temperature case}

Since the counter terms at finite temperature are the same as for the
vacuum case we can immediately write down the renormalized equations of
motion where we make use of the vacuum functions defined in (\ref{6.6}) and
the explicitly $T$-dependent finite part of the tadpole diagram:
\begin{equation}
\label{6.16}
\TP^{(T)}(M)=\frac{\ii \lambda}{2} \fint{l} 2 \pi n_{T}(l_{0})
\delta(l^{2}-M^{2}) = \frac{\ii \lambda}{4 \pi^{2}} \int_{M}^{\infty} \d \omega
\sqrt{\omega^{2}-M^{2}} n_{T}(\omega), 
\end{equation}
with the Bose-Einstein distribution function $n_{T}$.

\begin{figure}
\begin{minipage}{0.45\textwidth}
\centerline{\includegraphics[height=70mm]{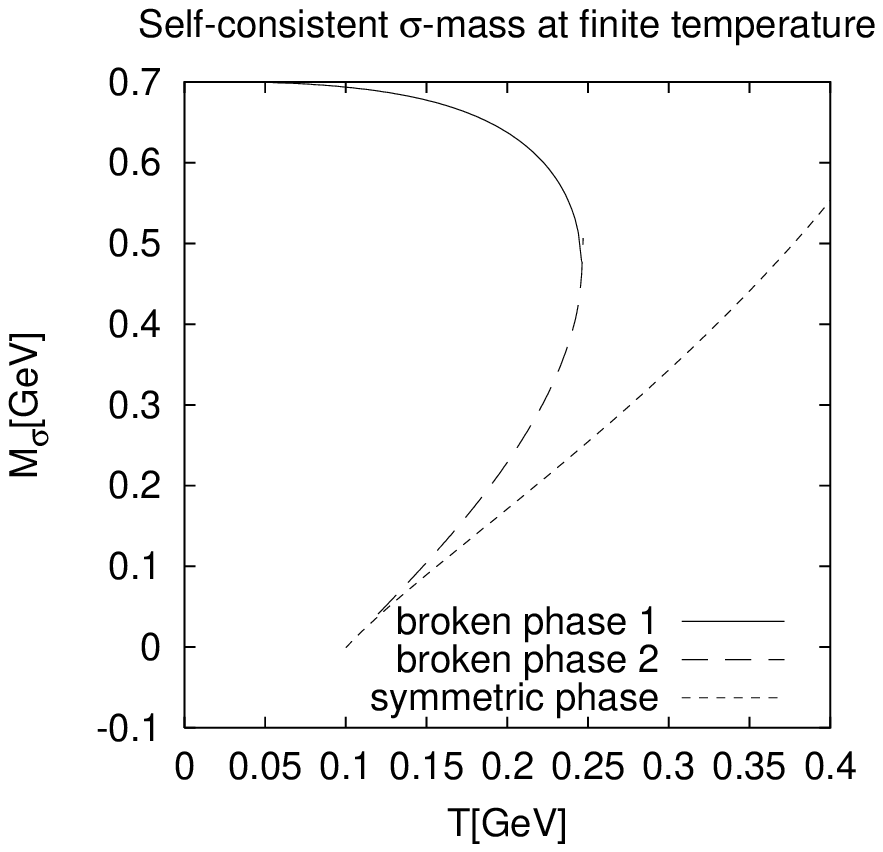}}
\end{minipage}
\begin{minipage}{0.45\textwidth}
\centerline{\includegraphics[height=70mm]{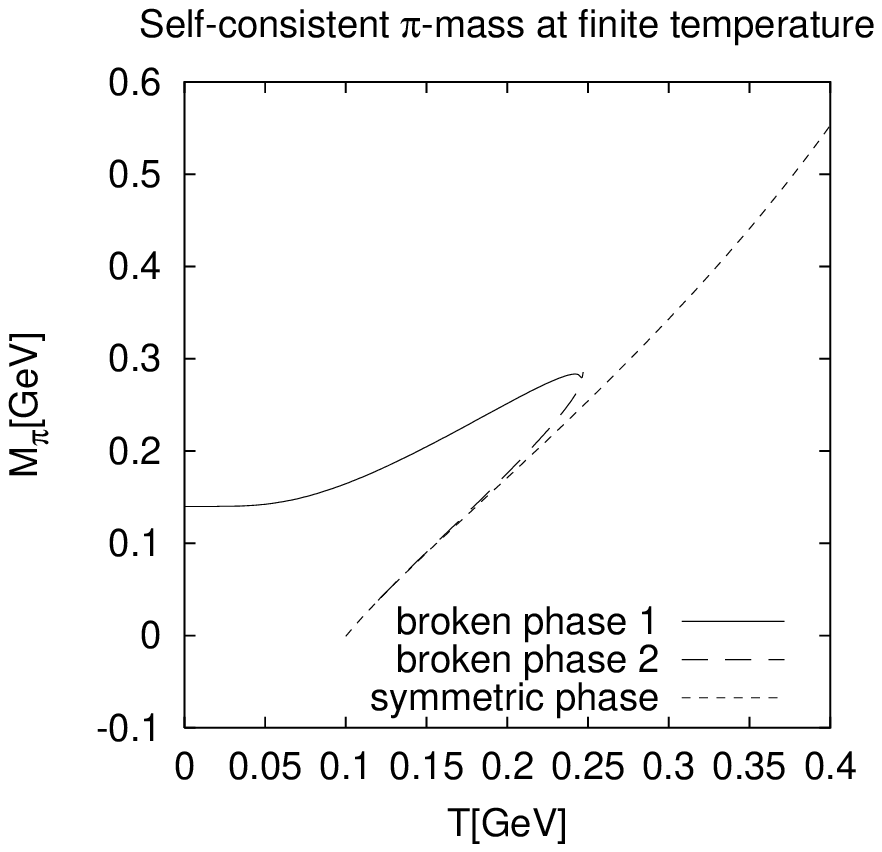}}
\end{minipage}
\caption{The solutions for the gap equations (\ref{6.17}).}
\label{fig.2}
\end{figure}

From the fact that the self-consistent $\{--\}$-propagator reads
\begin{equation}
\label{6.16b}
\begin{split}
G^{--}(p,M) &= \frac{1}{p^2-M^{2}+\ii \eta} - 2 \pi \ii n(p_0)
\delta_{\eta}(p^{2}-M^{2}), \\
\delta_{\eta}(x)&=\frac{1}{2 \pi \ii} \im \left (\frac{1}{x-\ii
    \eta}-\frac{1}{x+\ii \eta} \right)
\end{split}
\end{equation}
and that for our tadpole integral we are allowed take the limit $\eta
\rightarrow +0$ in the explicitly temperature dependent part we find the
renormalized gap equation at finite temperature
\begin{equation}
\label{6.17}
\begin{split}
  0 &= \varphi(M_{\parallel}^{2}-\lambda \varphi^{2}), \\
  M_{\perp}^{2} &= \TP_{\perp}(M_{\perp},M_{\parallel})+(N+1)
  \TP^{(T)}(M_{\perp}) +
  \TP^{(T)}(M_{\parallel}), \\
  M_{\parallel}^{2} &= \TP_{\parallel}(M_{\perp},M_{\parallel})+(N-1)
  \TP^{(T)}(M_{\perp}) + 3 \TP^{(T)}(M_{\parallel}),
\end{split}
\end{equation}
where $\TP_{\perp}$ and $\TP_{\parallel}$ are defined by
(\ref{6.6})\footnote{Note that although the self-consistent masses
  $M_{\perp}$ and $M_{\parallel}$ are temperature dependent the counter
  terms used to render $\TP_{\perp}$ and $\TP_{\parallel}$ finite are
  temperature-independent since the counter terms $\propto$
  $\Sigma_{\perp}$ and $\Sigma_{\parallel}$ are due to the subtraction of
  hidden vacuum-divergences of the four-point function as explained in
  detail in {\I}.}. The solutions of these gap-equations are shown in Fig.
\ref{fig.2} which shows clearly a first order phase transition behavior.
There exist two ``critical temperatures'': For $0<T<T_{c1}$ there exists
only one solution with $\varphi \neq 0$, for $T_{c1}<T<T_{c2}$ two
solutions with $\varphi \neq 0$ and the symmetric solution with $\varphi=0$
while for $T>T_{c2}$ the only solution is symmetric.
Also the equations of motion for the external self-energy (\ref{6.13})
remain formally the same but has to be read within the real-time
$\{-+\}$-matrix formalism. As cited in {\I} appendix A the analytic properties
of two-point functions lead to the conclusion that the retarded (and also
the advanced) propagators and self-energies decouple from the other degrees
of freedom, so that the simple algebraic properties are valid for them as
in the vacuum. We make use of the property
\begin{equation}
\label{6.18}
F^R(p)=\re F^{--}(p)+\ii \tanh \left(\frac{p_{0} \beta}{2} \right) \im
F^{--}(p),
\end{equation}
which holds true for any amputated two-point function.

In our special case we can take advantage of these analytic properties also
for the functions $\Lambda^{(4)}$, since effectively these are two-point
functions as well. Thus we can use the vacuum equations (\ref{6.11}) for
the retarded functions without changes.  Furthermore it is clear that these
functions do not contain any renormalization parts except the already
removed pure vacuum divergences.

Since the effective mass at $p=0$ is identical with the second derivative
of the effective potential, defined by
\begin{equation}
V_{\text{eff}}[\varphi] \delta^{(4)}(p) =
-\tilde{\Gamma}[\varphi]|_{\varphi=\text{const.}}
\end{equation}
its value should be $\geq 0$ for a stable solution, i.e., for a
minimum of the effective potential rather than a maximum which
provides an unstable ``tachyonic'' solution. The explicit calculation
shows that the solution, denoted by ``broken phase 2'' in Fig.
\ref{fig.2} are unstable. This shows that we find a phase transition
of first order, i.e., a discontinuity in the order parameter
$\varphi^{2}=M_{\parallel}^{2}/\lambda$ (see Fig. 1).

\begin{figure}
\begin{minipage}{0.45\textwidth}
\centerline{\includegraphics[width=\textwidth]{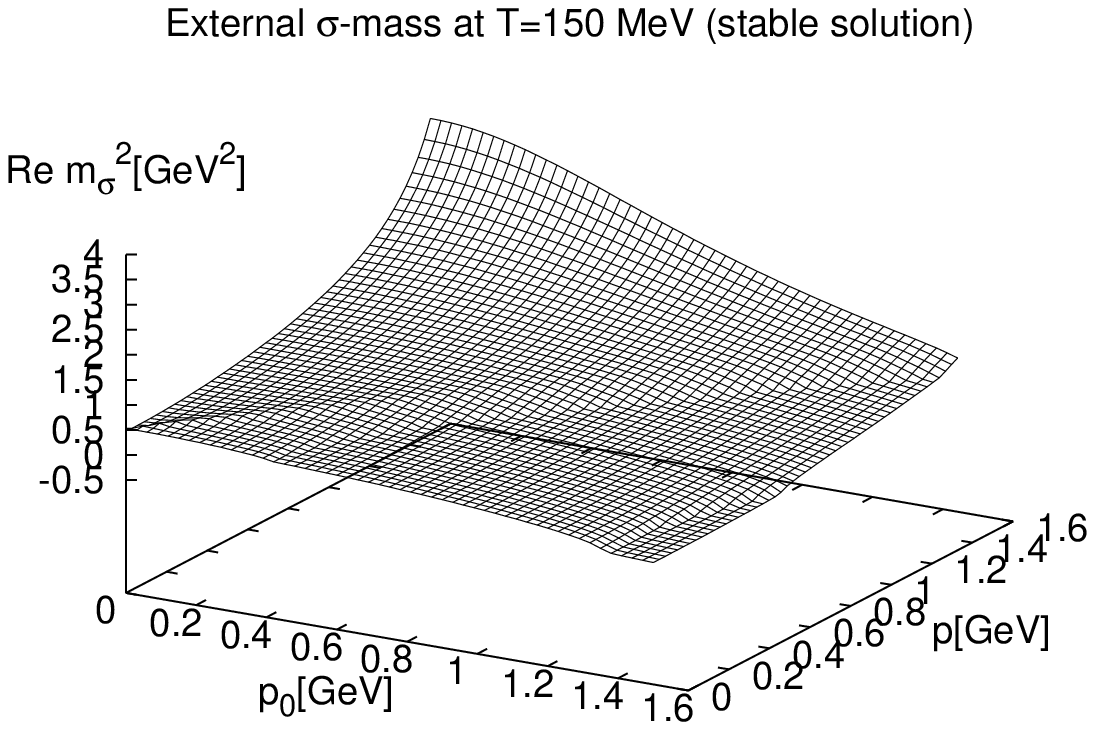}}
\vspace{5mm}
\end{minipage}
\begin{minipage}{0.45\textwidth}
\centerline{\includegraphics[width=\textwidth]{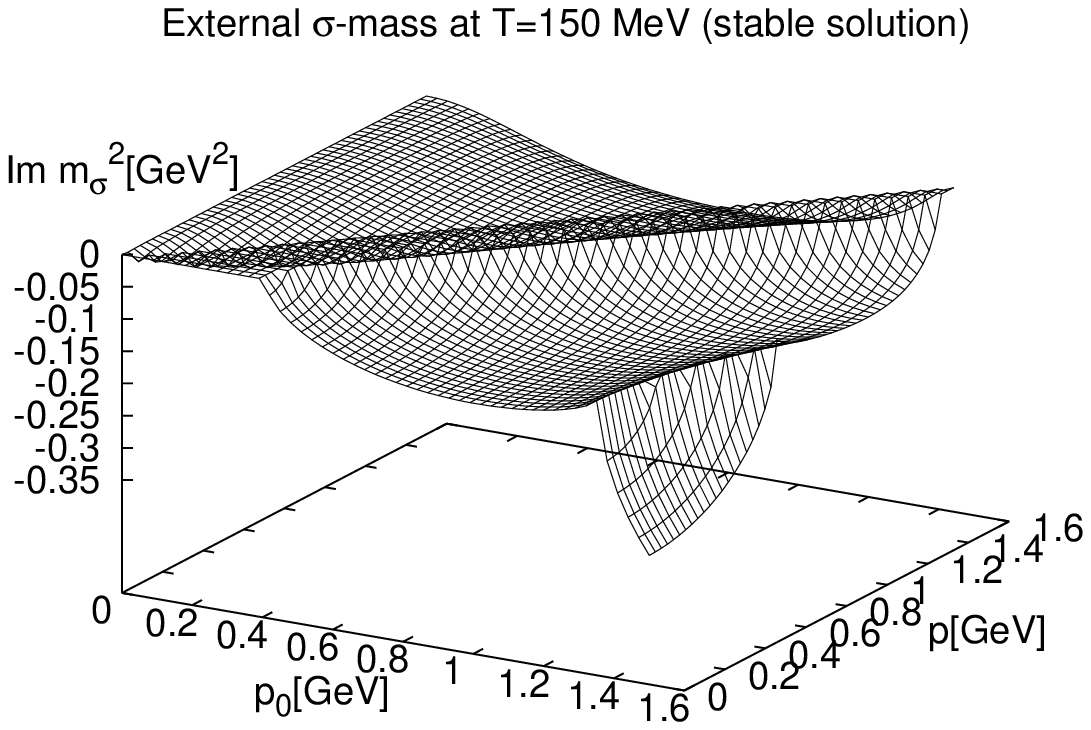}}
\vspace{5mm}
\end{minipage}
\begin{minipage}{0.45\textwidth}
\centerline{\includegraphics[width=\textwidth]{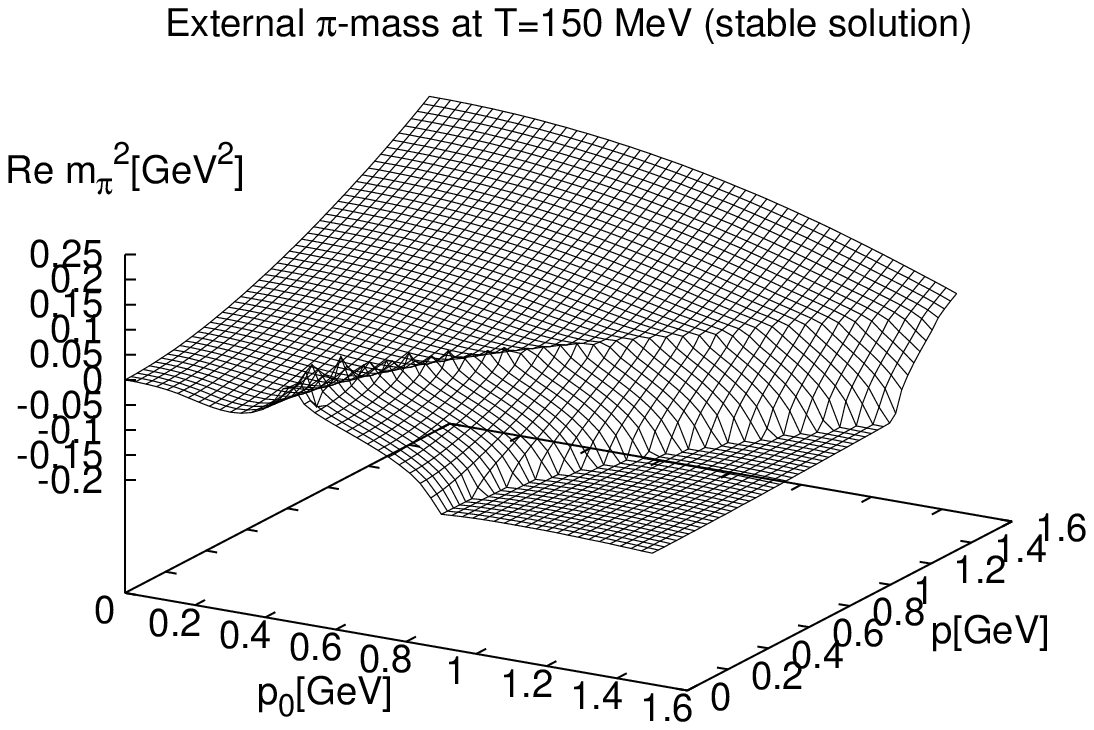}}
\end{minipage}
\begin{minipage}{0.45\textwidth}
\centerline{\includegraphics[width=\textwidth]{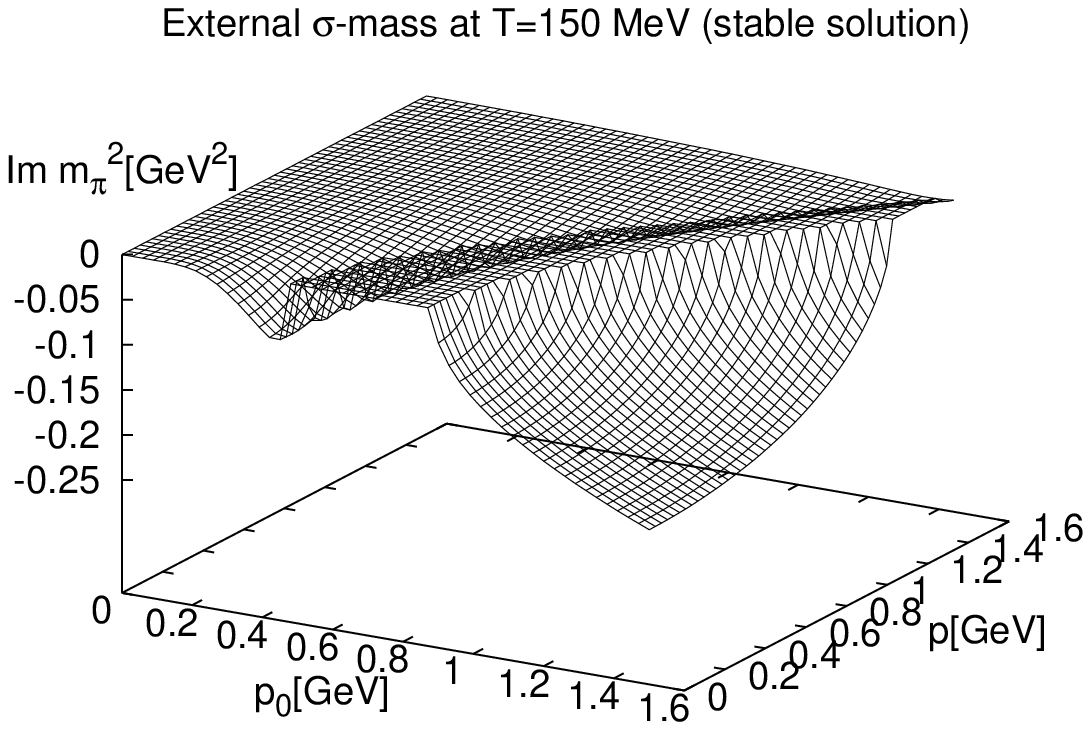}}
\end{minipage}
\centerline{\includegraphics[width=0.45\textwidth]{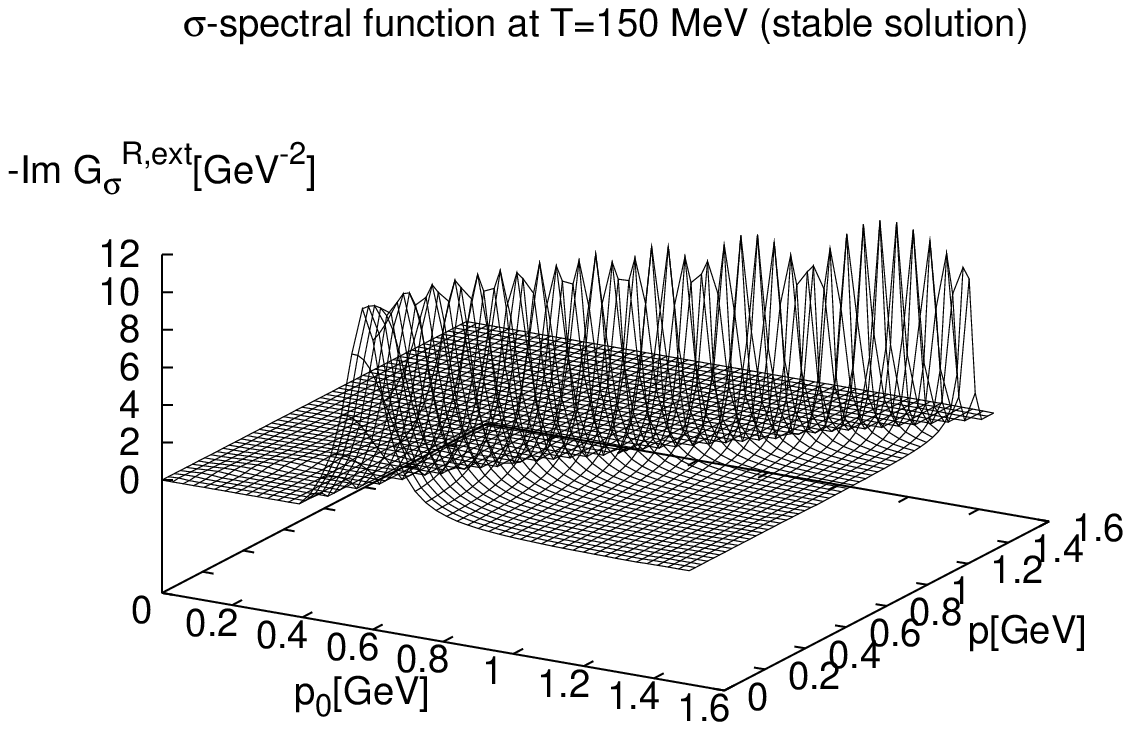}}
\caption{The effective external masses at a temperature of $150
  \text{MeV}$. The effective external $\pi$-mass indeed vanishes at
  $p_{0}=\vec{p}=0$ as predicted from Goldstone's theorem. The spectral
  function of the $\sigma$-meson shows that at high temperatures its
  strength becomes more peaked and the maximum shifted to lower momenta
  than at $T=0$.}
\label{fig.3}
\end{figure}

The effective masses for the stable spontaneously broken phase together
with the spectral function for the $\sigma$-meson are depicted in Fig.
\ref{fig.3}.

The calculation clearly shows the symmetry violations of the underlying
self-consistent Dyson approximation: Although the WTIs and Goldstone's
theorem are fulfilled for the external propagator remnants of their
violation by the internal propagators are present: The low-energy threshold
behavior of the ``$\sigma$-meson'' is not correct since the ``pion'' mass
within the internal propagator does not vanish. Also the phase transition
comes out to be of first instead of second order as it should be.

\subsection{Leading order large-N}

In the context of symmetries the large $N$ expansion scheme is of
particular importance. Here $N$ denotes the number of fields, e.g., in
SU($N$) or O($N$) theories. The counting depends on the type of theory and
is defined such that the classical action scales like $N$, c.f.
\cite{mot96,ahr02}. As unrestricted loops scale like $N$ this implies that
coupling and mean fields scale like
\begin{equation}
\label{scaling1/N}
\text{unrestricted loops} \propto N, \quad \lambda \propto \frac{1}{N},
\quad \vec{\varphi} \propto \sqrt{N}
\end{equation}
for the here considered O($N$) model in the Nambu-Goldstone phase. As the
WTIs concern the self-energies in the first place we shall first dicuss the
$1/N$ expansion at this level before we comment on it in the context of the
2PI-functional formalism.

\subsection*{$1/N$-expansion of the self-energy}
\label{subsec-1/N}
The $1/N$ counting scheme has the remarkable feature that the leading order
(LO) leads to zero order terms, i.e., $\propto (1/N)^0$, in the
self-energies.  This implies that iterative self-energy insertions of zero
order contribute in LO which amends the entire LO self-energy
self-consistently to be constructed within a corresponding Dyson
resummation.  Thus the resulting LO-propagator is created fully
self-consistently and at the same time accounts for {\em all} terms at LO.
The latter fact guarantees that for the LO self-energies the symmetries,
i.e., the corresponding WTIs, are fulfilled.

Since the above derived Hartree + RPA (external) self-energy, given by Eqs.
(\ref{6.3}) to (\ref{6.13}) indeed includes all zero order terms in
$1/N$-expansion, the leading order can simply by obtained by retaining the
according large $N$ limit terms
\begin{equation}
\label{6.13b1/N}
-\ii
 \Sigma^{\text{LO}}=\left\{\raisebox{1.6mm}
{\parbox{7.6cm}{\includegraphics{rpa-diag.eps}}}
 + \cdots\right\}^{\text{LO}}.
\end{equation}
Naturally this leads to counting factors different from those in the RPA
result (\ref{6.13b}) as different contractions of the
$(\vec{\varphi}\vec{\varphi})^2$ interaction term lead to different
$1/N$-orders (see, e.g., \cite{ahr02}). The counting further implies that
(a) solely loops in the Nambu-Goldstone modes survive and (b) therefore the
bubble sum only contributes to the $\sigma$-meson self-energy
$\Sigma_{\parallel}$.  The explicit result is
\begin{equation}
\label{6.20}
\begin{split}
  \Sigma_{\perp}^{\text{LO}}=&\frac{\ii N \lambda \mu^{2
      \epsilon}}{2}\feynint{l} G_{\perp}^{\text{LO}}(l),\quad
  \text{ with } (M_{\perp}^{\text{LO}})^{2} \varphi=0,  \\
  \Sigma_{\parallel}^{\text{LO}}=&\Sigma_{\perp}^{\text{LO}}
  +\frac{\varphi^2\lambda^2 N
      L_{\perp,\perp}^{\text{LO}}}{1-\lambda N L_{\perp,\perp}^{\text{LO}}}.
\end{split}
\end{equation}
Here $L_{\perp,\perp}^{\text{LO}}$ denotes the pion bubble loop, c.f.
(\ref{6.12}).  The result clearly shows that for the broken phase
$\varphi_l \neq 0$ the ``pion''-mass vanishes for the zero-order
propagator. The renormalization of Eq.  (\ref{6.20}) is done in the same
way as described above for the Hartree case and of course also the
renormalized ``pion''-mass vanishes for both, the vacuum and the
finite-temperature case. The WTI for the self-energy (\ref{6.20}) are
fulfilled by construction, if all terms of a given order are included, and
thus both the O($N$)-Noether current (\ref{3.7}) is conserved and the
Goldstone theorem is fulfilled. For the zero-order result (\ref{6.20}) all
this can be proven by mere inspection.

Please note that although the LO result (\ref{6.20}) can be generated from
a self-consistent Dyson resummation scheme, it is \emph{not
  $\Phi$-derivable}, since the pion self energy pieces corresponding to the
bubble series are of subleading order! Higher order approximations can even
not be expressed in terms of a self-consistent scheme, since from the Dyson
series the corresponding propagator contains all orders, while the terms
contributing to the self energy are of limited order!

\subsection*{$1/N$-expansion of the 2PI generating functional}

In recent times the $1/N$-expansion scheme also has been used to organize
truncation schemes for the here considered 2PI generating functional.
Thereby the counting results from the scaling rules (\ref{scaling1/N}),
however excluding the internal structure of the self-consistent propagators
from the counting in the diagrams of $\Phi$ (\ref{3n.5}), c.f. refs.
\cite{mot96,ahr02}. This procedure leads to a 2PI generating function which
is symmetric, a premise for the entire discussion in the papers.
  
In leading order $1/N$ one recovers the diagrams for the Hartree
approximation (\ref{6.1}) sect. \ref{sect-example} and the corresponding
Dyson resummation result, however with the counting factor arising from LO
in $1/N$. The internal self-energy resulting from the corresponding Dyson
resummation is given by the above result (\ref{6.13b1/N}), (\ref{6.20}),
however with the essential difference that the bubble contributions are
dropped. The reason is that the corresponding 2PI diagrams, c.f. Fig. 5 in
ref. \cite{ahr02}, are of subleading order. Nevertheless the pion mass is
still zero, since it is not affected from this difference, while the
sigma-meson self-energy is lacking the corresponding decay cuts and the
WTIs at the correlator level are indeed violated. The external self-energy
then exactly recovers the missing diagrams leading back to the result
(\ref{6.20}). In fact for the 2PI-$1/N$-expansion the Goldstone theorem can
only be assured for the so called mass-matrix which is identical to the
here considered external self-energy. A noteworthy side feature, though, is
that the LO approximation gives the correct $2^{\text{nd}}$-order phase
transition. For details see \cite{len99}.
  
As the main theme of this paper we explicitely see here that a symmetric
generating functional by itself does not guarantee that the solution of the
equations of motion, i.e., the Dyson equation, preserves the symmetry.
Rather the here presented functional scheme to construct \emph{external}
vertex functions is the minimal procedure to cure the corresponding
symmetry defects.

\section{Conclusion and Outlook}

In this paper we have analyzed Baym's $\Phi$-derivable approximations with
respect to their symmetry properties. It was shown that the self-consistent
mean field and propagator solutions of such approximations in general do not
fulfill the Ward-Takahashi identities of the full propagator as derived from
the usual 1PI generating functional, i.e., the effective action. The reason
for that lies in the fact that, although the expansion of the
\emph{2pI-functional} is done systematically in a symmetry-conserving ordering
scheme (for instance in powers of some coupling constant of a symmetric term
in the Lagrangian, the $\hbar$- or the $1/N$-expansion) the
\emph{solutions} of the equations of motion correspond to an incomplete
resummation to any order of the expansion parameter. In general not even
crossing symmetry is respected for the solutions beyond the order of the
expansion parameter.

Furthermore it could be shown, though, that for any such truncated Dyson
resummation scheme it is always possible to define a non-perturbative
1PI-effective action based on the self-consistent solution.  This
supplementary action indeed generates proper $n$-point vertex functions
which fulfill the whole hierarchy of Ward-Takahashi identities. These
\emph{external proper vertex functions} are implicitly determined by
\emph{closed} vertex equations of Bethe-Salpeter or higher order type,
where all ingredients are constructed from the self-consistent mean field
and the self-consistent propagator resulting from the underlying
$\Phi$-derivable Dyson scheme. The corresponding solutions exactly recover
the crossing symmetry and at the same time the symmetries of the original
classical action provided there are no intervening anomalies. The fact that
the $\Phi$-functional formalism was used to determine the self-consistent
mean fields and propagators does not only avoid double counting but also
ensures the consistency of counter terms, needed for renormalization of the
divergent integrals. This is valid for both, for the 2PI functional
$\Gamma[\vec{\varphi},G]$ and its equations of motion as well as for the
here discussed higher order vertex equations like the Bethe-Salpeter
equation resulting from the effective quantum action
$\tilde{\Gamma}[\vec{\varphi}]$. In the simple Hartree approximation the
symmetry repairing procedure leads to the well known Random Phase
Approximation (RPA). Similar features result for the leading order $1/N$
expansion of the 2PI functional.  For approximation levels with genuine
two-point self energies, i.e., for problems with damping, one arrives at
Bethe-Salpeter equations of ladder type, as they were considered, e.g., in
the context of Hard Thermal Loop (HTL) resummations or of photon production
\cite{knoll96}, which in semi-classical approximation leads to classical
transport equations.

Although the effective 1PI-action formally fulfills all the symmetry
properties of the original action it now suffers other defects essentially
resulting from the lack of self-consistency for the so constructed
\emph{external} multi-point functions. Their internal structure is
inherited from the self-consistent (e.g., Hartree) propagators: Since the
latter do not fulfill the Ward-Takahashi identities it is not guaranteed
that they show the correct dynamical behavior.  For instance in the case of
the here considered O($N$)-model in the spontaneously broken
Nambu-Goldstone phase without explicit symmetry breaking the transverse
(``pionic'') degrees of freedom have a non-zero mass in the self-consistent
propagator and thus all ``pion'' loops show a wrong dynamical behavior
(e.g., the vacuum ``$\sigma$'' spectral function shows a threshold at
$s=2M_{\perp}^{2}$ which is finite rather than zero!).  Another consequence
of this symmetry problem at the internal level is the wrong prediction of a
first order phase transition.

Both, the here considered 1PI and the 2PI effective actions
$\tilde{\Gamma}[\varphi]$ and $\Gamma[\varphi,G]$ have the same values
at the solution of the self-consistent solutions for the mean fields
and propagators and therefore represent a non-perturbative
approximation for the thermodynamical potential.  This implies that
the symmetry violation at the internal level may also leaves its traces in
this thermodynamical potential.

The derivation of approximation schemes that fulfill all symmetry
properties of the underlying classical action and at the same time are
fully self-consistent still remains as an open task.

In a forthcoming publication we shall discuss the symmetry properties for
the specially interesting case of local gauge symmetries.  Here the
background field method permits to establish an explicitly gauge-invariant
non-perturbative effective action on the basis of the 2PI-action
\cite{den96}, which leads to vertex functions that formally fulfill the
Ward-Takahashi identities. In this case, however, the symmetry violation by
the intermediate 2PI approximations causes even more serious complications,
namely the excitation of unphysical degrees of freedom for the gauge field
propagators. The latter fact may imply artifacts in the external proper
vertex functions which among others even may spoil thermodynamic
consistency due to the false number of degrees of freedom in the internal
lines. Here appropriate projection methods integrated into the Dyson scheme
as recently suggested by us \cite{vHK2001} may establish a suitable
work-around for such problems. For a more general recent review on
Schwinger Dyson equation approaches to non-abelian gauge theories, in
particular applied to the low-energy properties of QCD, see \cite{alk01}.

\section*{Acknowledgment}
We are grateful to D. Ahrensmeier, R. Baier, J. Berges, J. P.  Blaizot, P.
Danielewicz, B.  Friman, E. Iancu, Yu. Ivanov, M. Lutz, L. McLerran,
E. Mottola, R. Pisarski, and D.  Voskresensky for fruitful discussions
and suggestions at various stages of this work.

\appendix

\section{Calculation of the 2PI Functional}
\label{sect-baym-funct}

In order to give a precise meaning for the approximation schemes
considered we briefly derive the Feynman rules for the 2PI functional
in terms of path integrals. The main line of arguments follows
\cite{cjt74}. In order to perform a systematic loop expansion here we
explicitely shall keep track of $\hbar$ factors.

We shall restrict ourselves to theories with only scalar boson fields. The
generalization to other field contents is straight forward using the very
same functional integral techniques. Thus we assume the theory to be
defined in terms of a classical action functional
\begin{equation}
\label{2.1}
S[\phi]=\funcint{\Lag(\phi_{1})}{1},
\end{equation}
where $\phi$ denotes a multiplet of scalar fields. Here and in the
following we use the notation introduced in I: $\funcint{f(1,2,\ldots,n)}{1
  \ldots n}$ denotes the $d$-dimensional integral in the sense of
dimensional regularization with the $0$-component running along the
extended Schwinger-Keldysh path $\mathcal{C}$ (running from the initial
time $t_{i}$ to a final time $t_f$ and back along the real time axis and
then down to $t_{i}-\ii \beta$ parallel to the imaginary time axis and at
the end taken $t_{i} \rightarrow -\infty$ and $t_{f} \rightarrow +
\infty$). Here this contour integral also implies the appropriate sums over
charge-space indices.

The generating functional with local and bilocal sources is defined by
\begin{equation}
\label{2.2}
Z[J,B]=N \int \D \phi \exp \left [ \frac{\ii}{\hbar} S[\phi]+
    \frac{\ii}{\hbar} \funcint{J_{1}
    \phi_{1}}{1} + 
\frac{\ii}{2 \hbar} \funcint{B_{12} \phi_{1} \phi_{2}}{12}
\right],
\end{equation}
where $N$ is a indefinite normalization constant which will be chosen such
that for the temperature $T \rightarrow 0$ and $J=B=0$ the functional is
normalized to $1$.

Shifting the integration variable of the path integral by an arbitrary
field $\varphi$ one obtains
\begin{equation}
\label{2.3}
\begin{split}
  Z[J,B]= & N \exp \left [ \frac{\ii}{\hbar} S[\varphi]+ \frac{\ii}{\hbar}
    \funcint{J_{1} \varphi_{1}}{1} + \frac{\ii}{2 \hbar}
    \funcint{B_{12} \varphi_{1} \varphi_{2}}{12} \right]
  \times \\ &\times \underbrace{\int \D \phi' \exp \left [ \frac{\ii}{2
        \hbar} \funcint{(G^{-1})_{12} \phi_{1}' \phi_{2}'}{12}
      + \frac{\ii}{\hbar} S_{I}[\phi',\varphi] + \frac{\ii}{\hbar}
      \funcint{J_{1}' \phi_{1}'}{1} \right]}_{Z_{1}[J']},
\end{split} 
\end{equation}
where we introduced the following abbreviations
\begin{equation}
\label{2.4}
\begin{split}
  (G^{-1})_{12} &= \frac{\delta^{2} S[\varphi]}{\delta \varphi_{1}
    \delta \varphi_{2}} + B_{12}:=(\mathscr{D}^{-1})_{12} + B_{12}, \\
  S_{I}[\phi',\varphi] &= S[\phi'+\varphi]-S[\varphi]-\funcint{
    \funcd{S[\varphi]}{\varphi_{1}} \phi_{1}'}{1} - \frac{1}{2}
  \funcint{\frac{\delta^2 S[\varphi]}{\delta \varphi_{1}
      \delta \varphi_{2}} \phi_{1}' \phi_{2}'}{12}, \\
  J_{1}' &= J_{1} + \funcd{S[\varphi]}{\varphi_{1}} + \funcint{B_{11'}
    \varphi_{1'}}{1'}.
\end{split}
\end{equation}
Now we want to chose $J$ and $B$ such that $\varphi$ and $G$ are the exact
mean field and the exact propagator respectively. These are defined with
help of $W[J,B]=-\ii \hbar \ln Z[J,B]$ via
\begin{equation}
\label{2.5}
\varphi_{1} = \funcd{W[J,B]}{J_{1}}, \; \ii \hbar G_{12} =
2 \funcd{W[J,B]}{B_{12}} - \varphi_{1} \varphi_{2}.
\end{equation}
From the first condition we derive immediately that we have to chose
$J'=J_{0}'$ such that
\begin{equation}
\label{2.6}
\left . \funcd{Z_{1}[J']}{J'} \right |_{J'=J_{0}'} \stackrel{!}{=}0.
\end{equation}
In this way the functional $Z_{1}[J_{0}']$ defines completely $Z[J,B]$ and
can be calculated approximately with well-known standard techniques (see,
e.g., \cite{itz80}). An example will be given in Sect. \ref{sect-on-funct}.
In this way any expansion with respect to the number of loops (powers of
$\hbar$) or with respect to the coupling constant defines a
$\Phi$-derivable approximation. Of course any other expansion scheme, known
from usual perturbation theory, is feasible. E.g., recently in
\cite{berges01b} and \cite{ahr02} investigations of the Large-$N$-expansion
description for Baym's functional were undertaken.

For sake of completeness we give the derivation of Baym's
$\Phi$-functional. To that end we only have to perform the
$\hbar$-expansion up to first order, i.e., up to one-loop order in terms of
diagrams. For this task it is convenient to introduce the new action
\begin{equation}
\label{2.7}
\tilde{S}[\phi';J']=\frac{1}{2} \funcint{(G^{-1})_{12} \phi_{1}'
  \phi_{2}'}{12} + S_{I}[\phi',\varphi] + \funcint{J_{1}' 
  \phi_{1}'}{1}
\end{equation}
so that the functional $Z_{1}$, defined in (\ref{2.3}), reads
\begin{equation}
\label{2.8}
Z_{1}[J']=N \int \D \phi \exp \left [\frac{\ii}{\hbar} \tilde{S}[\phi';J']
\right] = \exp \left ( \frac{\ii}{\hbar} W_{1}[J'] \right ).
\end{equation}
To obtain the $\hbar$-expansion we have to expand the functional integral
around the solution of the classical field equations $\varphi_{0}'$, which
is the stationary point of the classical action:
\begin{equation}
\label{2.9}
\funcd{\tilde{S}[\varphi_{0}';J']}{\varphi_{01}'}=0
\end{equation} 
and substitute $\sqrt{\hbar} \phi''=\phi'-\varphi_{0}'$:
\begin{equation}
\label{2.10}
Z_{1}[J']=N \exp \left \{ \frac{\ii}{\hbar} \tilde{S}[\varphi_{0}',J']
\right \}
\int \D \phi'' \exp \left [\frac{\ii}{2} \funcint{(\mathscr{G}^{-1})_{12}
    \phi_{1}'' \phi_{2}''}{12} +
  \frac{\ii}{\hbar} \tilde{S}_{I}[\sqrt{\hbar} \phi'',\varphi_{0}'] \right]
\end{equation}
with the definitions
\begin{equation}
\label{2.11}
\begin{split}
  (\mathscr{G}^{-1})_{1j,2k} &= \frac{\delta^{2}
    \tilde{S}[\varphi_{0}';J']}{\delta
    \varphi_{01} \delta \varphi_{02}}, \\
  \tilde{S}_{I}[\sqrt{\hbar} \phi'',\varphi_{0}'] &=
  \tilde{S}[\varphi_{0}'+\sqrt{\hbar} \phi'';J']-\tilde{S}[\sqrt{\hbar}
  \varphi_{0}';J'] - \frac{\hbar}{2} \funcint{(\mathscr{G}^{-1})_{12}
    \phi_{1}'' \phi_{2}''}{12}.
\end{split}
\end{equation}
Note that $(\mathscr{G}^{-1})_{12}$ and $\tilde{S}_{I}$ both depend on $J'$
only implicitly via $\varphi_{0}'$ and that
$\tilde{S}_{I}^{(k)}[\sqrt{\hbar} \varphi'',\varphi_{0}]=O[\hbar^{k/2}]$
where $\tilde{S}_{I}^{(k)}$ denotes the monomial to order $\phi''{}^{k}$.
By definition only the terms with $k \geq 3$ are different from $0$. The
most general renormalizable theory has only $k=3$ and $k=4$ contributions
in $\tilde{S}_{I}$ and we shall restrict ourselves to this case.

Now it is easy to extract the one-loop contribution (i.e., the
$O(\hbar)$-contribution) to the generating functional $W'$ for connected
diagrams:
\begin{equation}
\label{2n.12}
\begin{split}
  W_{1}[J'] &= -\ii \hbar \left [ \ln N + \frac{\ii}{\hbar}
    \tilde{S}[\varphi_{0}',J'] + \ln \int \D \phi'' \exp \left [
      \frac{\ii}{2} \mathscr{G}^{-1}_{12}
      \phi_{1}'' \phi_{2}'' \right] + W_{2}[J'] \right ] \\
  &= \tilde{S}[\varphi_{0}',J'] + \frac{\ii \hbar}{2} \Tr \ln (
  \mathscr{G}^{-1}_{12}/M^{2} ) +W_{2}'[J'].
\end{split} 
\end{equation}
Herein we have introduced an arbitrary mass scale $M$ to avoid
dimensionful quantities within the logarithm which takes account of the
overall normalization of $Z$ which is irrelevant for any physical quantity
derived from it. Now according to (\ref{2.6}) we have to chose $J'=J_{0}'$
such that $\varphi'=\delta W_{1}/\delta J'=0$, and we have
\begin{equation}
\label{2n.13}
\varphi_{1}':=\funcd{\tilde{S}[\varphi_{0}',J']}{J_{1}'} + O(\hbar),
\end{equation}
so that we can substitute $\varphi'$ instead of $\varphi_{0}'$ in
(\ref{2n.12}) leading only to a modification of the functional $W_{2}'$ to
order $O(\hbar^{2})$ while the $O(\hbar)$-part remains unchanged.

According to (\ref{2.6}) we have to chose $J'=J_{0}'$ such that
$\varphi'=0$ to obtain the original $W$-functional. We also note that for
this choice $\mathscr{G}$ becomes $G$ according to (\ref{2.3}):
\begin{equation}
\label{2n.14}
W[J,B] = -\ii \hbar \ln Z[J,B]= S[\varphi]+ \funcint{J_{1} \varphi_{1}}{1}
+ \frac{1}{2}
\funcint{B_{12} \varphi_{1} \varphi_{2}}{12} + W_{1}[J_{0}'].
\end{equation}
Now we define the \emph{2PI effective action} by the double Legendre
transformation of $W$ with respect to $J$ and $B$. Using (\ref{2.5}) this
leads to
\begin{equation}
\label{2n.15}
\begin{split}
  \Gamma[\varphi,G] &= W[J,B]-\funcint{J_{1} \varphi_{1}}{1} - \frac{1}{2}
  \funcint{(\varphi_{1} \varphi_{2} + \ii \hbar G_{12}) B_{12}}{12} \\
  &= S[\varphi] + \frac{\ii \hbar}{2} \Tr \ln (G^{-1}/M^{2})
  +\frac{\ii \hbar}{2}
  \funcint{(\mathscr{D}^{-1})_{12}(G_{12}-\mathscr{D}_{12})}{12} +
  \Phi[\varphi,G].
\end{split}
\end{equation}
From the derivation we note that $\Phi[\varphi,G]=O(\hbar^{2})$, i.e., in
the language of diagrams it contains only diagrams with at least two loops.
The main difference to the usual \emph{1PI effective action} is that the
lines appearing in the diagrams symbolize full propagators $\ii G$ rather
than perturbative ones.

The equations of motion are now given by the fact that we like to study the
theory for vanishing auxiliary sources $J$ and $B$. From the Legendre
transformation (\ref{2n.14}) we can immediately express this in terms of
the functional $\Gamma$:
\begin{equation}
\label{2n.16}
\begin{split}
  \funcd{\Gamma[\varphi,G]}{\varphi} &=-J_{1} -\funcint{B_{12}
    \varphi_{12}}{2}
  \stackrel{!}{=} 0 \\
  \funcd{\Gamma[\varphi,G]}{G} &= -\frac{\ii \hbar}{2} B_{12}
  \stackrel{!}{=} 0.
\end{split}
\end{equation}
Using the $\hbar$-expansion (\ref{2n.15}) the last line reads:
\begin{equation}
\label{2n.17}
\mathscr{D}_{12}^{-1}-G_{12}^{-1} = \frac{2 \ii}{\hbar}
\funcd{\Phi[\varphi,G]}{G_{12}} := \Sigma_{12}. 
\end{equation}
It is clear that $\Sigma_{12}$ is the \emph{exact self-energy} expressed in
terms of the \emph{exact connected propagator} $G$ and the \emph{exact mean
  field} $\varphi$ and thus (\ref{2n.17}) is the full self-consistent
\emph{Dyson equation}. This implies that no propagator line in the diagrams
must contain any self-energy insertion, because these lines denote already
the full propagator. Thus $\Phi[\varphi,G]$ consists of all closed diagrams
with point vertices from the action $S_{I}[\varphi,\phi]$ with at least two
loops\footnote{Note that $\phi$ denotes the ``quantum field'', integrated
  over within the path integral and $\varphi$ the ``classical background
  field''. $S_{I}$ consists of the sum over all monomials with at least
  three quantum fields.}, which have the additional property that it is
impossible to disjoin them by cutting only one or two lines, i.e., all
diagrams contained on $\Phi$ must be \emph{two-particle irreducible} (2PI).
This must hold true because taking the derivative of $\Phi$ with respect to
$G$ according to (\ref{2n.17}) defines the full proper self-energy which
must be 1PI and no propagator line should contain any self-energy
insertion. Now taking this derivatives diagrammatically means to open any
propagator line of all diagrams contained in $\Phi$ and taking the sum of
the so obtained skeleton self-energy diagrams. This indeed implies the 2PI
property for the diagrams defining $\Phi$ since otherwise one could disjoin
a diagram by cutting two lines, leading to a 1PR diagram contributing to
the self-energy which by definition cannot be contained in the diagram
expansion of the self-energy.

In Sect. \ref{sect-on-funct} on the analytical example for the
O($N$)-linear $\sigma$-model up to order $\hbar^{2}$ we see that the double
Legendre transformation (\ref{2n.15}) indeed leads to 2PI diagrams for
$\Phi[\varphi,G]$. In this section also an example of diagrammatical
derivations for the $\Phi$-functional for the same theory is given.

\section{Symmetries of the classical action}
\label{app-a}

For sake of completeness we summarize the derivation of Noether's theorem
\cite{noe18} for symmetries of a classical action functional.  For sake of
simplicity we shall restrict ourselves again to a multiplet of scalar
fields. The extension to more general cases is straight forward.

We investigate the behavior of the classical action functional
$S[\vec{\phi}]$ under a general infinitesimal transformation of the form
\begin{equation}
\label{a.1}
x'{}^{\mu}=x^{\mu}+\delta x^{\mu}, \;
\vec{\phi}'(x')=\vec{\phi}(x)+\delta \vec{\phi}(x),
\end{equation}
where $\delta \vec{\phi}$ may depend on both, the fields and the space-time
argument. The action functional is defined to be symmetric under the
transformation (\ref{a.1}) if its variation
\begin{equation}
\label{a.2}
\delta S[\vec{\phi}]=\int \d^{d} x' \Lag(\vec{\phi}',\partial_{\mu}'
\vec{\phi}',x') - \int \d^{d} x \Lag(\vec{\phi},\partial_{\mu}
\vec{\phi},x) \equiv 0.
\end{equation}
vanishes \emph{identically}, i.e., without any restrictions on the fields.
To derive explicit conditions for $\Lag$ to fulfill the symmetry condition
(\ref{a.2}) we have to rewrite the first integral in terms of $x$, where we
have to take into account the Jacobian of the volume element. In linear
order of the variation we have
\begin{equation}
\label{a.3}
\det \left (\frac{\partial x'{}^{\mu}}{\partial x^{\nu}} \right) =
1+\partial_{\mu} \delta x^{\mu}.
\end{equation}
The transformation and differentiation with respect to the space-time
arguments does not commute:
\begin{equation}
\label{a.4}
\delta (\partial_{\mu} \phi):=\partial_{\mu}' \phi'(x')-\partial_{\mu}
\phi = \partial_{\mu}(\delta \phi)-(\partial_{\mu} \delta x^{\nu})
\partial_{\nu} \phi.
\end{equation} 
After some algebraic manipulations we find
\begin{equation}
\label{a.5}
\delta S[\phi]=\int \d^n x \funcd{S[\phi]}{\phi} \left [ \delta \phi -
  (\partial_{\nu} \phi) \delta x^{\nu} \right] \equiv 0.
\end{equation}
Let now $\delta \eta^{a}$ be the independent parameters of the Lie group
acting on the fields and the space-time variables, i.e., (\ref{a.1}) reads
\begin{equation}
\label{a.6}
\delta \phi(x)=\tau_{a}(x,\phi) \delta \eta^a, \; \delta x^{\mu} =
-\tilde{\tau}_{a}^{\mu}(x,\phi) \delta \eta^a.
\end{equation}
This means that the symmetry condition (\ref{a.5}) reads
\begin{equation}
\label{a.7}
\int \d^{n} x \funcd{S[\phi]}{\phi(x)} \left \{ \tau_{a}(x,\phi)+
  [\partial_{\nu} \phi(x)] \tilde{\tau}_{a}^{\nu}(x,\phi) \right \}
\delta \eta^a \equiv 0. 
\end{equation}
Since this must hold for \emph{any field configuration} for which the
action is well defined and the $\delta \eta^a$ are independent generators
of the group operation for each $a$ there must exist a current $j^a$, the
\emph{Noether currents} corresponding to the symmetry group, such that
\begin{equation}
\label{a.8}
 \funcd{S[\phi]}{\phi(x)} \left \{ \tau_{a}(x,\phi)+
  [\partial_{\nu} \phi(x)] \tilde{\tau}_{a}^{\nu}(x,\phi) \right \} =
\partial_{\mu} j_{a}^{\mu}.
\end{equation}
Now the classical field equations of motion are given by the stationarity
of the action. Thus \emph{for the solutions of the equations of motion the
  Noether-currents} are conserved.

To find the explicit expression for the Noether-currents we go back to
(\ref{a.5}) and express it in terms of the Lagrangian. After some
calculations we obtain
\begin{equation}
\label{a.9}
\delta \eta^{a} \partial_{\mu} j_{a}^{\mu}= \partial_{\mu} \left [
  \left ( (\partial_{\nu} \phi) \frac{\partial
      \Lag}{\partial(\partial_{\mu} \phi)} - \delta_{\nu}^{\mu} \Lag
  \right) \delta x^{\nu} - \frac{\partial
    \Lag}{\partial(\partial_{\mu} \phi)} \delta \phi \right] + \delta
\Lag + \Lag \partial_{\mu} \delta x^{\mu}. 
\end{equation}
This means that there must exist local functionals
$\Omega_{a}^{\mu}(\phi,x)$ such that
\begin{equation}
\label{a.10}
\delta \Lag+\Lag \partial_{\mu} \delta x^{\mu}=\partial_{\mu}
\Omega_{a}^{\mu} \delta \eta^{a},
\end{equation}
which leads to the desired explicit expression for the conserved Noether
current:
\begin{equation}
\label{a.11}
\delta \eta^{a} j_{a}^{\mu} =  \left ( (\partial_{\nu} \phi) \frac{\partial
      \Lag}{\partial(\partial_{\mu} \phi)} - \delta_{\nu}^{\mu} \Lag
  \right) \delta x^{\nu} - \frac{\partial
    \Lag}{\partial(\partial_{\mu} \phi)} \delta \phi +\Omega_{a}^{\mu}
  \delta \eta^{a}.
\end{equation}
It should be noted that within special relativity only the total conserved
quantities are physical observables, not the local currents themselves.
Those follow from the continuity equation $\partial_{\mu} j^{\mu}=0$ by
integration over a space-like closed hypersurface of space-time. Taking the
special hypersurface given in a Lorentz reference frame by $x^0=t_0$ and
$x^{0}=t_{1}$ we find
\begin{equation}
\label{a.12}
Q(t_{1})-Q(t_{0})=\int_{t_{0}}^{t_{1}} \d x^{0} \int_{\R^{3}} \d^{3}
\vec{x} \; \partial_{0} j^{0}(\vec{x})=\int_{t_{0}}^{t_{1}} \int_{\R^{3}}
\d^{3} \vec{x} \; \text{div} \; \vec{j} =0, 
\end{equation}
where we have made use of the continuity equation. Eq. (\ref{a.12}) tells
us that the \emph{total Noether charge}
\begin{equation}
\label{a.13}
Q(t)=\int_{\R^{3}} \d^{3} \vec{x} j^{0} = \text{const.}
\end{equation}
For sake of completeness it should be mentioned that the currents
(\ref{a.11}) are not determined uniquely by the symmetries since changing
it according to
\begin{equation}
\label{a.14}
(j')^{\mu}=\partial_{\rho} \omega^{\rho \mu} \text{ with }\omega^{\rho \mu}
= -\omega^{\mu \rho} 
\end{equation}
does not change the Noether charge (\ref{a.13}), and for $j'$ the
continuity equation holds true as well as for $j$. This notion is important
especially in the context of particles of higher spin where the freedom of
choice of the Noether current, in (\ref{a.14}) parameterized by the
antisymmetric tensor $\omega^{\rho \mu}$, can be used to give gauge
invariant definitions of the energy momentum tensor. In general its
canonical version
\begin{equation}
\label{a.15}
\Theta^{\mu \nu}= \partial^{\mu} \frac{\partial
  \Lag}{\partial(\partial_{\nu} \phi)} -g^{\mu \nu} \Lag
\end{equation} 
is not gauge invariant. It is clear from (\ref{a.11}) that (\ref{a.15}) is
the ``Noether current'' of the symmetry of the physical laws against
space-time translations which leads to energy-momentum conservation:
\begin{equation}
\label{a.16}
\frac{\d}{\d t} \int \d^{3} \vec{x} \; \Theta^{0 \nu}(x)=0.
\end{equation}

\section{Some Feynman integrals}
\label{app-b}

In this appendix we give three dimensionally regularized Feynman integrals,
needed for the application to the tadpole approximation and the
corresponding RPA-summed external self-energies. The techniques to obtain
them can be found in standard textbooks, for instance in \cite{ram89}. As
usual we set $d=4-2 \epsilon$ for the space-time dimension in the sense of
dimensional regularization, $\mu$ for the regularization scale; $\gamma
\approx 0.577$ stands for the Euler-Mascheroni constant.

The first integral is the vacuum tadpole with a free propagator for a mass
$m$:
\begin{equation}
\label{b.1}
\TP_{1}(m^{2})=\ii \fint{l} \frac{\mu^{2\epsilon}}{l^{2}-m^{2}+\ii \eta} =
\frac{m^{2}}{16 \pi^2} \left [-\frac{1}{\epsilon} -1 + \gamma + \ln \left (
    \frac{m^{2}}{4 \pi \mu^{2}}
  \right )\right ] + O(\epsilon).
\end{equation}
The next integral is used to renormalize the divergent four-point
sub-diagrams contained in the tadpole approximation as a hidden divergence:
\begin{equation}
\label{b.2}
\begin{split}
  \TP_{2}(m^{2}) &= \ii \fint{l} \frac{\mu^{2 \epsilon}}{(l^{2}-m^{2}+\ii
    \eta)^{2}} = \frac{1}{2m} \partial_{m} \TP_{1}(m^{2}) \\
  &= \frac{1}{16 \pi^{2}} \left [ -\frac{1}{\epsilon}+\gamma + \ln \left (
      \frac{m^{2}}{4 \pi \mu^{2}} \right )\right ] + O(\epsilon).
\end{split}
\end{equation}
Further for the calculation of the external self-energy we need the
two-point function
\begin{equation}
\label{b.3}
\begin{split}
  L_{m_{1},m_{2}}(p^{2}) &= \ii \fint{l} \frac{mu^{2
      \epsilon}}{(l^{2}-m_{1}^{2}+\ii \eta)[(l-p)^{2}-m_{2}^{2} + \ii
    \eta]} \\
  &= \frac{1}{16 \pi^{2}} \Bigg \{ -\frac{1}{\epsilon} -2 + \gamma +
  \frac{\lambda(p^{2},m_{1}^{2},m_{2}^{2})}{p^{2}} \\
  & \quad \quad \quad \times \Bigg [ \artanh \left
    (\frac{m_{1}^{2}-m_{2}^{2}+p^{2}}{\lambda(p^{2},m_{1}^{2},m_{2}^{2})}
  \right) + \artanh \left
    (\frac{m_{2}^{2}-m_{1}^{2}+p^{2}}{\lambda(p^{2},m_{1}^{2},m_{2}^{2})} +
  \right) \Bigg ]  \\ & \quad \quad \quad + \frac{m_{1}^{2}-m_{2}^{2}}{p^{2}}
  \ln \left ( \frac{m_{1}}{m_{2}} \right) + \ln \left ( \frac{m_{1}
      m_{2}}{4 \pi \mu^{2}} \right ) \Bigg \} + O(\epsilon),
\end{split}
\end{equation}
where the K{\"a}ll{\`e}n function reads
\begin{equation}
\label{b.4}
\lambda(p^{2},m_{1}^{2},m_{2}^{2}) =
\sqrt{[p^{2}-(m_{1}+m_{2})^{2}][p^{2}-(m_{1}-m_{2})^{2}]}.
\end{equation}
For the proof of Goldstone's theorem for the external propagator we need
this function at $p=0$ which can be expressed with help of the already
defined function $\TP_{1}$:
\begin{equation}
\label{b.5}
L_{m_{1},m_{2}}(0) = \frac{1}{m_{1}^{2}-m_{2}^{2}} \left [\TP_{1}(m_{1}^{2})
  -  \TP_{1}(m_{2}^{2}) \right ]
\end{equation}
Their expressions for equal masses read
\begin{equation}
\label{b.6}
\begin{split}
L_{m,m}(p) = \frac{1}{16 \pi^2} \Bigg [ & -\frac{1}{\epsilon}-2 + \gamma +
\ln \left (\frac{m^{2}}{4 \pi \mu^{2}} \right ) \\
 & \quad \quad \quad + 2 \frac{\lambda(p^{2},m^{2},m^{2})}{p^{2}} \artanh
 \left (  \frac{p^{2}}{\lambda(p^{2},m^{2},m^{2})} \right) \Bigg ]
\end{split}
\end{equation}
and for $p=0$ this expression becomes
\begin{equation}
\label{b.7}
L_{m,m}(0) = \frac{1}{2m} \partial_{m} \TP_{1}(m^{2})=\TP_{2}(m^{2}).
\end{equation}

\begin{flushleft}

\end{flushleft}

\end{document}